\documentclass[aps,twocolumn,floatfix,groupedaddress,nofootinbib]{revtex4}
\usepackage{csquotes}
\usepackage{graphicx,color}
\usepackage{float}
\usepackage[caption=false]{subfig}
\usepackage{amsmath}
\usepackage{amsmath}
\usepackage{multirow}
\usepackage{dsfont}
\usepackage[shortlabels]{enumitem}
\usepackage{epstopdf}

\begin{document}

\title{Testing the isomorph invariance of the bridge functions of Yukawa one-component plasmas. II. Short range}
\author{\vspace*{-2.50mm}F. Lucco Castello$^{1}$, P. Tolias$^{1}$ and J. C. Dyre$^{2}$}
\affiliation{$^{1}$Space and Plasma Physics, Royal Institute of Technology, Stockholm, SE-100 44, Sweden\\
             $^{2}$Glass and Time, IMFUFA, Roskilde University, Roskilde, DK-4000, Denmark}
\begin{abstract}
\noindent It has been conjectured that bridge functions remain nearly invariant along phase diagram lines of constant excess entropy for the class of R-simple liquids. In the companion paper, this hypothesis has been confirmed for Yukawa bridge functions outside the correlation void. In order to complete the testing of the invariance ansatz, Yukawa bridge functions are here computed inside the correlation void with the cavity distribution method and input from ultra-long molecular dynamics simulations featuring a tagged particle pair. A general methodology is developed for the design of the tagged pair interaction potential that leads to the acquisition of uniform statistics. An extrapolation technique is developed to determine the bridge function value at zero separation. The effect of different sources of errors is quantified. Yukawa bridge functions are demonstrated to be nearly isomorph invariant also in the short range. Generalization to all R-simple systems and practical implications are discussed.
\end{abstract}
\maketitle\vspace*{-10.50mm}

\section{Introduction}\label{sec:intro}

\noindent \emph{Yukawa one-component plasmas (YOCP)} are model systems that consist of charged point particles which are immersed in a polarizable charge-neutralizing background. The particle interactions follow the Yukawa (or screened Coulomb) pair potential
\begin{equation}
u(r)=\frac{Q^2}{r}\exp{\left(-\frac{r}{\lambda}\right)}\,,
\end{equation}
where $Q$ is the particle charge and $\lambda$ is the screening length that is determined by the polarizable background medium. It is convenient to specify the thermodynamic state points of the YOCP in terms of the following independent dimensionless variables, the coupling parameter $\Gamma$ and the screening parameter $\kappa$ defined by\,\cite{dustrev1,dustrev2,dustrev3,dustrev4}
\begin{equation}
\Gamma=\beta\frac{Q^2}{d}\,,\qquad\kappa=\frac{d}{\lambda}\,.
\end{equation}
Here, $\beta=1/(k_{\mathrm{B}}T)$ where $k_{\mathrm{B}}$ is the Boltzmann constant, $T$ the temperature and $d=(4\pi{n}/3)^{-1/3}$ for the Wigner-Seitz radius with\,$n$\,the particle (number) density. In normalized units, the Yukawa potential reads as
\begin{equation}
\beta{u}(x)=\frac{\Gamma}{x}\exp{\left(-\kappa{x}\right)}\,,
\end{equation}
where $x=r/d$. In the limit of a non-polarizable (or rigid) background, $\lambda\to\infty$ or $\kappa\to0$, particles interact with the unscreened Coulomb potential and this model system is known as the one-component plasma (OCP)\,\cite{OCPrevs1,OCPrevs2,OCPrevs3}. Important realizations of YOCP systems are complex plasmas\,\cite{dustexp1,dustexp2}, charged colloids\,\cite{dustexp3}, ultra-cold neutral plasmas\,\cite{dustexp4}, and perhaps even warm dense matter\,\cite{dustexp5}.

A recent computational study has revealed that dense YOCP liquids belong to the rather broad class of many-body systems that have been coined as Roskilde-simple or R-simple\,\cite{isoYOCPg,isogene1,isogene2,isogene3,isogene4,isogene5}. R-simple systems are in practice identified from the strong correlations that emerge between their virial ($W$) and their potential energy ($U$) constant-volume thermal equilibrium fluctuations\,\cite{isogene1}. Such systems possess isomorphic curves or simply isomorphs, \emph{i.e.} phase diagram curves of constant excess entropy, along which a large set of structural and dynamic properties are approximately invariant when expressed in properly reduced units\,\cite{isogene2,isogene3,isogene4,isogene5}. In particular, it has been demonstrated that the YOCP exhibits exceptionally high $W-U$ correlation coefficients for an extended part of the fluid phase that covers its entire dense liquid region\,\cite{isoYOCPg}.

A novel integral equation theory approach has been formulated for R-simple systems with isomorph invariance as the basic building block\,\cite{ourwork1}. The isomorph-based empirically modified hypernetted-chain (IEMHNC) approach is based on the conjecture that, when expressed in properly reduced distance units, the bridge functions are invariant along isomorphic curves\,\cite{ourwork1}. In other words, it is assumed that the bridge functions of R-simple systems depend exclusively on the reduced distance and on the reduced excess entropy. The IEMHNC approximation has been applied to Yukawa and bi-Yukawa liquids exhibiting a remarkable agreement with computer simulations in terms of thermodynamic and structural properties at a relatively low computational cost\,\cite{ourwork1,ourwork2,ourwork3}.

The main objective of the present investigation is to provide a rigorous test of the validity of the underlying ansatz of bridge function invariance for dense YOCP liquids. For this reason, Yukawa bridge functions have been indirectly extracted from computer simulations for multiple thermodynamic state points along different phase diagram lines of constant excess entropy. Owing to the emergence of the correlation void at short distances, different computational methods must be used in the intermediate or long range and in the short range. In the first article\,\cite{accompan}, hereafter referred to as Paper I, the intermediate and long range of Yukawa bridge functions was computed along different isomorphic lines with the Ornstein-Zernike inversion method and \enquote{exact} input from ultra-accurate standard canonical NVT molecular dynamics (MD) simulations. In the present article, the short range of Yukawa bridge functions is computed along different isomorphic lines with the cavity distribution method and \enquote{exact} input from ultra-long specially designed canonical NVT MD simulations.

Before proceeding to the contents of the present work, we shall first summarize the main results of Paper I\,\cite{accompan}. Sixteen Yukawa state points, equally distributed amongst four isomorphs, were specified by the application of three isomorph tracing techniques, namely the direct isomorph check\,\cite{isogene6,isogene7,isogene8}, the small step method\,\cite{isogene9,isogen10} and the analytical method\,\cite{isogen11}. These $16$ YOCP state points were selected in a manner that roughly covers the entire dense liquid regime of interest. In what follows, the four YOCP isomorphs will be uniquely identified by the coupling parameter of their $\kappa=0$ OCP member $\Gamma_{\mathrm{ISO}}^{\mathrm{OCP}}$ or the nearly constant ratio $\Gamma/\Gamma_{\mathrm{m}}$\,\cite{isogene7,isomana1}, where
\begin{equation}
\Gamma_{\mathrm{m}}(\kappa){e}^{-\alpha\kappa}\left[1+\alpha\kappa+\frac{1}{2}(\alpha\kappa)^2\right]=\Gamma_{\mathrm{m}}^{\mathrm{OCP}}\,,\label{analytical_melting}
\end{equation}
is a semi-empirical description of the YOCP melting line\,\cite{isomana2,isomana3} which accurately follows the near-exact MD data\,\cite{isomana4,isomana5}. In the above, $\alpha=\Delta/d=(4\pi/3)^{1/3}$ denotes the ratio between the mean-cubic inter-particle distance and the Wigner-Seitz radius, while $\Gamma_{\mathrm{m}}^{\mathrm{OCP}}=171.8$ is the OCP coupling parameter at melting\,\cite{isomana4}. The YOCP bridge functions were computed outside the correlation void by means of the Ornstein-Zernike inversion method and radial distribution function input from carefully designed MD simulations. Rigorous uncertainty propagation analysis was guided by a detailed investigation of the bridge function sensitivity to periodic and aperiodic multiplicative perturbations in radial distribution functions. The effects of statistical, grid, finite-size, tail and isomorphic errors were quantified. Within the long and intermediate range, the YOCP bridge functions were concluded to be only approximately isomorph invariant, since the small deviations observed between different members of the same isomorph were demonstrated to exceed the overall uncertainties.

This paper is organized as follows. In section \ref{BridgeExtraction}, a recapitulation of the integral equation theory of liquids, the Ornstein-Zernike inversion method and the cavity distribution method is provided that emphasizes the necessity of uniform particle statistics and highlights the central role of the correlation void. In section \ref{CavityMD}, the theoretical basis of the cavity distribution method is outlined, the design of the tagged pair potential for the acquisition of uniform statistics is described (focusing on its decomposition in windowing and biasing components) and the numerical implementation of the method is detailed. In section \ref{CavityBridge}, the short range bridge functions are computed for all the $16$ YOCP state points and an extrapolation method is developed that allows for the determination of the bridge function value at zero separation. In section\,\ref{UncertaintyPropagation}, the effect of different types of errors in the short range bridge function is quantified for all the $16$ YOCP state points of interest. In section \ref{UncertaintyBridge}, the corrected short range bridge functions featuring error bars due to propagating uncertainties are presented and the degree of isomorph invariance is examined. Finally, in section\,\ref{BridgeFull}, the bridge function results in the full range are discussed in their totality together with implications and future work.

\section{Integral equation theory and bridge function extraction methods}\label{BridgeExtraction}

\noindent In the case of a one-component pair-interacting isotropic system, the integral equation theory of liquids consists of the Ornstein-Zernike (OZ) equation\,\cite{Hansenbo,liquidbo,theorybo,bridger1,bridger2}
\begin{equation}
h(r)=c(r)+n\int c(r')h(|\boldsymbol{r}-\boldsymbol{r}'|)d^3r'\,,\label{eq:theory_oz}
\end{equation}
combined with the formally exact non-linear closure condition\,\cite{Hansenbo,liquidbo,theorybo,bridger1,bridger2}
\begin{equation}
g(r)=\exp\left[-\beta u(r)+h(r)-c(r)+B(r)\right]\,,\label{eq:theory_oz_closure}
\end{equation}
with $g(r)$ the radial distribution function, $h(r)=g(r)-1$ the total correlation function, $c(r)$ the direct correlation function, $B(r)$ the bridge function. Auxiliary static two-particle correlation functions of relevance are the indirect correlation function $\gamma(r)=h(r)-c(r)$, the potential of mean force $\beta{w}(r)=-\ln{[g(r)]}$, the screening potential $\beta{H}(r)=\beta{u(r)}-\beta{w}(r)$ and the cavity distribution function $y(r)=g(r)\exp{[\beta{u}(r)]}$. A formally exact expression for the bridge function is required to close the above system of equations.

The radial distribution function has an intuitive physical interpretation being equal to the probability density of finding a particle at a distance from a reference particle relative to the probability density for an ideal gas\,\cite{theorybo}. The cavity distribution function also has a physical interpretation being equal to the radial distribution function for a pair of tagged particles whose mutual interaction is suppressed that are dissolved at infinite dilution in a system where all other interactions remain the same\,\cite{Hansenbo,liquidbo}. The latter function remains continuous even if the interaction potential is discontinuous or diverges and acquires large but finite values near the origin $r=0$\,\cite{theorybo}. Finally, the potential of mean force has a physical interpretation, since the opposite of its gradient is equal to the force exerted on one member of a particle pair that is held at fixed positions, after averaging over all possible positions of the remaining particles\,\cite{Hansenbo}.

In contrast to the radial distribution function and the cavity distribution function, the bridge function neither possesses a microscopic representation (\emph{i.e.} it cannot be expressed as the ensemble average of a function that depends on the instantaneous particle positions) nor a physical interpretation (\emph{e.g.} in terms of a probability density). Within the framework of diagrammatic analysis, bridge functions are graphically represented by highly connected diagrams, that contain neither nodal points nor articulation points and their root points do not form articulation pairs. This implies that their evaluation is very cumbersome\,\cite{bridger3}. In fact, the bridge function is formally defined through the f-bond expansion $B(r)=\sum_{i=2}^{\infty}d_i(r;T)n^{i}$ where the coefficients $d_i(r;T)$ are given by a number of multi-dimensional integrals whose kernels are products that involve Mayer functions $f(r)=\exp{[-\beta{u}(r)]}-1$, or through the h-bond expansion $B(r)=\sum_{i=2}^{\infty}b_i(r;n,T)n^{i}$ where the coefficients $b_i(r;n,T)$ are given by a number of multi-dimensional integrals whose kernels are products that involve total correlation functions $h(r)$\,\cite{bridger4,bridger5}. As the order of the coefficients increases, the number of their integral constituents rises dramatically and the complexity of each kernel increases rapidly\,\cite{bridger6,bridger7}. More importantly, both bond expansions are known to converge very slowly already at moderate densities\,\cite{bridger3,bridger5,bridger6,bridger7}.

As a consequence, the direct extraction of bridge functions from computer simulations is not possible and the computation of bridge functions through their formal definition is a formidable task even with modern computational means. Nevertheless, bridge functions can be calculated with input from computer simulations by exploiting the fact that radial distribution functions, as well as cavity distribution functions, can be extracted from computer simulations and by taking advantage of the exact expressions of integral equation theory.

\begin{figure}
	\centering
	\includegraphics[width=3.05in]{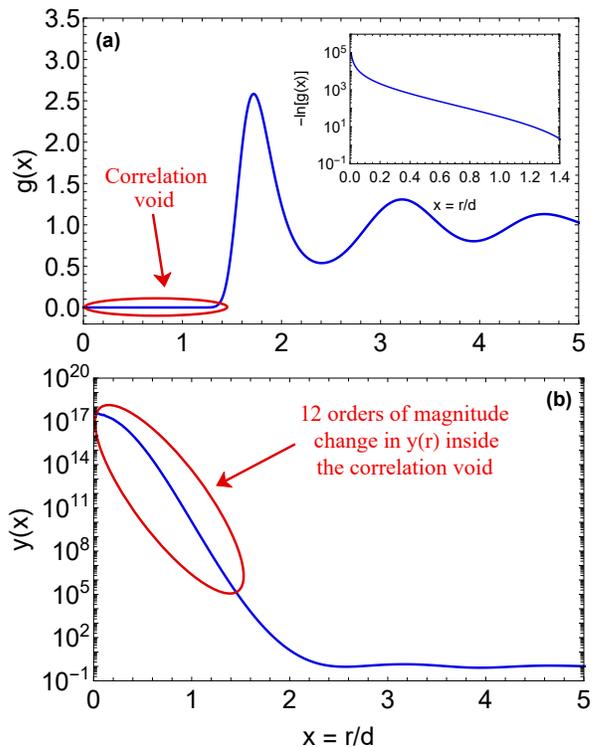}
	\caption{Characteristics of the radial distribution function (a), the cavity distribution function (b) for dense simple liquids. MD results for the YOCP state point $\Gamma=708.517$, $\kappa=2.5$ ($\Gamma_{\mathrm{ISO}}^{\mathrm{OCP}}=160$). The correlation void, $\displaystyle\mathrm{arg}_{r}\{g(r)\simeq0\}$, is accompanied by a rapid drop in the pair particle statistics (see the inset) and a rapid increase in the cavity statistics. However, the cavity statistics are strongly non-uniform at short distances, as confirmed by the twelve orders of magnitude difference between the cavity distribution function value at the edge of the correlation void, $r\lesssim1.4d$, and at the origin, $r=0$.}\label{fig:introduction}
\end{figure}

In the \emph{Ornstein-Zernike inversion method}, the radial distribution function is extracted from MD or from MC simulations. The direct correlation function is then computed from the OZ equation and the bridge function is computed from the closure condition,
\begin{equation}
B(r)=\ln{[g(r)]}-g(r)+\beta{u}(r)+c(r)+1\,\label{bridgeequationfull}\,.
\end{equation}
The extraction of radial distribution functions with the histogram method fails at short distances that lie within the so-called correlation void, where particle pair encounters are ultra rare especially for dense systems and, thus, the collected statistics are poor even in the course of very long simulations. The correlation void can be loosely defined as $\displaystyle\mathrm{arg}_{r}\{g(r)\ll1\}$ or $\displaystyle\mathrm{arg}_{r}\{g(r)\simeq0\}$ and its exact range depends on the thermodynamic state point of interest, see figure \ref{fig:introduction}(a) for an illustration.  For the YOCP simulations reported herein, the correlation void roughly corresponds to $r\lesssim1.4d$. From the structure of the OZ equation, it is straightforward to see that the direct correlation function is insensitive to the exact small values of the radial distribution function inside the correlation void. Therefore, the direct correlation function can be accurately computed across the whole range with the OZ inversion method. On the other hand, due to the $\ln{[g(r)]}$ term which appears in the closure condition, it is evident that the bridge function is very sensitive to the exact small value of the radial distribution function inside the correlation void. Therefore, the bridge function can only be reliably computed in the intermediate and long ranges, \emph{i.e.} outside the correlation void, with the OZ inversion method. This computational method was used in Paper I, where the main challenge identified was the acquisition of large statistical samples, since the bridge function in the intermediate and long range has a very strong sensitivity to the radial distribution function\,\cite{accompan}.

In the \emph{cavity distribution method}, the cavity distribution function is directly extracted from MD or MC simulations. With knowledge of the direct correlation function from the OZ inversion method, the bridge function is then computed from the closure condition that, within the correlation void, reads
\begin{equation}
B(r)\simeq\ln{[y(r)]}+c(r)+1\,.\label{bridgeequation}
\end{equation}
As illustrated in figure \ref{fig:introduction}(b), the cavity distribution function acquires very high values within the correlation void, which suggests that large global sample statistics can be obtained. However, the values of the cavity distribution function increase by many orders of magnitude from the edge of the correlation void up to the origin $r=0$, which implies that uniform sample statistics are rather impossible to acquire owing to the fact that the very localized sub-interval close to $r=0$ will always be over-sampled. Thus, a type of umbrella sampling technique needs to be followed that inserts a known bias which homogenizes the statistics within the entire correlation void. This turns out to be the main challenge in the implementation of the cavity distribution method. Finally, we emphasize that the cavity distribution method needs to be complemented with the OZ inversion method for the computation of the bridge function due to the presence of the direct correlation function in the closure equation.

\section{The cavity simulations}\label{CavityMD}

\subsection{Theoretical basis of the simulation method}\label{SubCavityThe}

\noindent In the cavity simulations, two tagged particles are interacting with a specially designed pair potential that allows them to explore all the nearly-forbidden distances within the correlation void. The potential of mean force that is exerted on the tagged particles still originates from the remaining $N-2$ particles. The tagged pair potential adds an externally controlled bias that allows the sampling of the ultra-rare pair configurations within the correlation void. The known statistical bias can then be removed, so that the statistical weights that correspond to the actual static correlations are ultimately extracted. The cavity simulation method was originally developed by Torrie \& Patey\,\cite{cavitym1} for hard sphere potentials and non-interacting tagged particles and later generalized by Llano-Restrepo \& Chapman\,\cite{cavitym2} to arbitrary interaction potentials and arbitrary tagged pair potentials.

The \emph{targeted system} is a simple one-component liquid which consists of $N$ particles that interact with the pair potential $u(r)$. The \emph{simulated system} is a binary simple liquid mixture which consists of $N-2$ type A particles that interact with the pair potential $u(r)$ and two type B particles (tagged as 1,2) that interact with the pair potential $u(r)$ with the type A particles, but interact via a pair potential $\psi(r)$ with each other. This system's static correlation functions will be denoted with the script \enquote{sim}. The total potential energies of these two systems are connected via $U_{\mathrm{sim}}(\boldsymbol{R})=U(\boldsymbol{R})+\psi(\boldsymbol{r}_1,\boldsymbol{r}_2)-u(\boldsymbol{r}_1,\boldsymbol{r}_2)$ with $\boldsymbol{R}$ the N-particle configuration given by the collective position vector $(\boldsymbol{r}_1,...,\boldsymbol{r}_N)$. The reduced two-particle density in the targeted system and reduced two-particle density of the tagged particles in the simulated system read as
\begin{align*}
&n_2(\boldsymbol{r}_1,\boldsymbol{r}_2)=N(N-1)\frac{\int\exp{\left[-\beta{U}(\boldsymbol{R})\right]}d^3r_3...d^3r_{N}}{\int\exp{\left[-\beta{U}(\boldsymbol{R})\right]}d^3r_1...d^3r_{N}}\,,\\
&n_2^{\mathrm{sim}}(\boldsymbol{r}_1,\boldsymbol{r}_2)=2\frac{\int\exp{\left[-\beta{U}_{\mathrm{sim}}(\boldsymbol{R})\right]}d^3r_3...d^3r_{N}}{\int_{V_{\mathrm{c}}}\exp{\left[-\beta{U}_{\mathrm{sim}}(\boldsymbol{R})\right]}d^3r_1...d^3r_{N}}\,,
\end{align*}
in the canonical ensemble. The introduction of the constraint volume $V_{\mathrm{c}}$ in the configuration integral of the simulated system originates from the fact that, in order to enhance statistics, the tagged particle pair is not allowed to explore the whole configuration space. The numerical pre-factors stem from particle indistinguishability and integrand symmetry with respect to particle label interchange. Combining the above, we acquire
\begin{align*}
\frac{n_2^{\mathrm{sim}}(\boldsymbol{r}_1,\boldsymbol{r}_2)}{n_2(\boldsymbol{r}_1,\boldsymbol{r}_2)}=\frac{e^{-\beta{\psi}(\boldsymbol{r}_1,\boldsymbol{r}_2)}}{e^{-\beta{u}(\boldsymbol{r}_1,\boldsymbol{r}_2)}}\frac{2\int{e}^{-\beta{U}(\boldsymbol{R})}d^3r^{N}}{N(N-1)\int_{V_{\mathrm{c}}}{e}^{-\beta{U}_{\mathrm{sim}}(\boldsymbol{R})}d^3r^{N}}
\end{align*}
where we introduced the notation\,$d^3r_1...d^3r_{N}=d^3r^{N}$ for brevity. We point out that the second factor is independent of the tagged particle positions $\boldsymbol{r}_1,\boldsymbol{r}_2$. In case the constraint volume $V_{\mathrm{c}}$ is smaller than the primary cell volume of the simulation, this factor cannot be evaluated in the course of the simulation\,\cite{cavitym2}. In what follows, it will be denoted with $1/C$. By introducing the radial distribution functions of both systems $g(\boldsymbol{r}_1,\boldsymbol{r}_2)=n_2(\boldsymbol{r}_1,\boldsymbol{r}_2)/n^2$, exploiting the isotropy of the interaction potentials and introducing the cavity distribution function of the targeted system $y(r_{12})=g(r_{12})\exp{[\beta{u}(r_{12})]}$ with $r_{12}=|\boldsymbol{r}_1-\boldsymbol{r}_2|$, we finally obtain
\begin{equation}
y(r)=C\exp{[\beta{\psi}(r)]}g_{\mathrm{sim}}^{12}(r)\,.\label{cavityequation}
\end{equation}
after setting $r=r_{12}$ and $g_{\mathrm{sim}}^{12}(r)=g_{\mathrm{sim}}(r_{12})$ in order to emphasize that the pair correlations of tagged particles should be sampled. Owing to the continuity of the cavity distribution function, the unknown constant $C$ should be determined by matching with known $y(r)$ values either outside the correlation void (obtained by OZ inversion) or at the origin (obtained by Widom's expansion\,\cite{cavitym3}). The former option is preferable not only because it is characterized by smaller statistical errors but also because such simulations are readily available\,\cite{accompan}.

Overall, in cavity simulations, the radial distribution function of the two tagged particles $g_{\mathrm{sim}}^{12}(r)$ is extracted with the histogram method which leads to the determination of the short range distance dependence of the cavity distribution function $y(r)$ of the targeted system, see Eq.(\ref{cavityequation}). The unknown proportionality constant $C$ is then quantified by a matching procedure with available intermediate range $y(r)$ results at the upper edge of the correlation void which leads to the determination of the short range cavity distribution function $y(r)$ of the real system, see Eq.(\ref{cavityequation}) again. Finally, the short range bridge function $B(r)$ can be determined from Eq.(\ref{bridgeequation}) with knowledge of the short range $\ln{[y(r)]}$ and the full range $c(r)$, the latter reliably obtained from the OZ inversion method\,\cite{accompan}. The cavity simulation method is straightforward, but there are inherent difficulties that are connected with the specification of the tagged pair potential $\psi(r)$. In fact, as we shall see in the sections that follow, multiple preliminary cavity simulations will be required in order to make an informed guess for $\psi(r)$ that is accurate enough for reliable bridge functions to be computed.

\subsection{Windowing and biasing components of the tagged pair potential}\label{SubCavityBias}

\noindent The tagged pair potential $\psi(r)$ should ensure that the entire correlation void is sampled as uniformly as possible by the tagged particle pair. Otherwise, large statistical errors will emerge in the poorly sampled annular rings. It is evident that the tagged pair potential $\psi(r)$ should strongly depend on the thermodynamic state point. This can be illustrated by assuming $\psi(r)=0$\,\cite{cavitym2}. In this case, the non-interacting tagged particles will tend to remain overlapped at most sampled configurations for dense systems close to crystallization and they will tend to remain at large separations for gas-like systems of low density.

The acquisition of uniform statistics along the correlation void is a complicated task, since it requires an accurate guess of the potential of mean force $-\ln{[g(r)]}$ that in turn requires an accurate guess of the short range bridge function $B(r)$ that is our unknown. It is more practical to separate the correlation void into a number of successive overlapping windows ($I_n=[b_n,c_n]$ with $c_{n}>b_{n+1}$) by imposing hard-constraints in the tagged particle motion. In fact, the potentials of mean force can be guessed more accurately with an iterative computational scheme at shorter intervals wherein the relative variations in $g(r)$ are far less dramatic, while the overlap is required in order to determine the unknown proportionality constant $C$ of Eq.(\ref{cavityequation}) by means of sequential matching\,\cite{cavitym4,cavitym5}. On the other hand, the hard constraints can also be viewed as flexible bond lengths and be implemented through the tagged pair potential itself\,\cite{cavitym6}. Overall, the tagged pair potential $\psi(r)$ is decomposed into a windowing component $\chi(r)$ that confines the tagged particles within given finite intervals and a biasing component $\phi(r)$ that uniformly samples each windowing interval, \emph{i.e.}
\begin{equation}
\psi_n(r)=\chi_n(r)+\phi_n(r)\,,\label{potentialdecompositionequation}
\end{equation}
where $n$ is the number of successive overlapping windows.

It is preferable that the \emph{windowing component} of the tagged pair potential does not affect the correlation sampling within each window, implying that $d\chi(r_{n})/dr\simeq0,\,\forall{r}_n\in{I_n}$. It is also essential that the windowing component is steep enough at the vicinity of each window's edges, so that the tagged particles cannot traverse the restriction zone even in the course of long simulations. An infinite potential well appears to be the most straightforward mathematical implementation, since it describes a tagged particle that is free to move within small distances from the other particle surrounded by impenetrable barriers. Owing to the involvement of infinities, it becomes necessary to treat the impulsive elastic collisions, implying that the practical implementation becomes quite involved. As a result, the smooth finite potential well that is generated by the sum of two error functions was preferred. In normalized units $\beta\chi(x)$ with $x=r/d$, we have
\begin{equation}
\beta\chi_n(x)=a_{1}\left\{\mathrm{erf}\left[a_{2}(a_{3n}-x)\right]+\mathrm{erf}\left[a_{2}(a_{4n}-x)\right]\right\}\label{windowingequation}
\end{equation}
where the state variable dependent coefficient $a_{1}(n,T)$ controls the depth of the well, the constant coefficient $a_{2}$ controls the steepness of the well, the window dependent coefficients $(a_{3n},a_{4n})$ control the extent and position of the well. The coefficients $(a_2,a_{3n},a_{4n})$ should be selected such that $\chi(x)\simeq\mathrm{constant}$ within each window, while the coefficients $(a_{1},a_2)$ should be selected such that tagged particles do not escape the windowing interval. It is worth pointing out that very steep realizations might turn out to be problematic and counter-intuitively lead to the loss of tagged particle confinement due to the insufficient MD time steps. After consideration of the above rough guidelines, the exact values of these coefficients are determined by trial and error.

In each window, the \emph{biasing component} of the tagged pair potential, which should counteract the absent pair repulsion and the potential of mean force, is initially determined through a series of short cavity simulations with the following iterative procedure, that is based on Eq.(\ref{cavityequation}) and the multiplicative identity of exponentials: \textbf{(a)} The tagged particles are assumed to be non-interacting, a cavity simulation is performed, $g_{\mathrm{sim}}^{12}(x)$ is extracted and its logarithm is interpolated with a Gaussian function $G_1(x)$. \textbf{(b)} The biasing component is assumed to be given by $\beta\phi(x)=G_1(x)$, another cavity simulation is performed, $g_{\mathrm{sim}}^{12}(x)$ is extracted and its logarithm is interpolated with a Gaussian function $G_2(x)$. \textbf{(c)} The biasing component is then assumed to be given by $\beta\phi(x)=G_1(x)+G_2(x)$ and the procedure is repeated resulting in an additional Gaussian function. \textbf{(d)} The iterative procedure is terminated, when the targeted interval is judged to be sufficiently sampled. In this manner, the biasing component is approximated by a series of Gaussian functions, \emph{i.e.}
\begin{equation}
\beta\phi_n(x)=\sum_{i=1}^{k}d_{1,i,n}\exp{\left[-d_{2,i,n}(x-d_{3,i,n})^2\right]}\,,\label{biasingequation}
\end{equation}
where $k$ is the number of iterations and the coefficients $(d_{1,i,n}\,,d_{2,i,n},\,d_{3,i,n})$ are determined by least square fits. The iteration scheme proved to be robust and leads to analytical approximations for the biasing potential. Its major disadvantage is concerned with the fact that $\beta\phi_n(x)$ is not necessarily monotonic, which can lead to tagged particle trapping in the course of very long simulations. Such trapping would cause non-uniform sampling and would induce spurious correlations between samples collected at different time steps, thus generating uncertainties that would be hard to quantify. The following remedy proved to be very successful: \textbf{(a)} Long MD cavity simulations are carried out at each interval using Eqs.(\ref{potentialdecompositionequation},\ref{windowingequation},\ref{biasingequation}), the sub-interval cavity distribution functions are matched and the cavity distribution function $y(r)$ is extracted in the whole domain. \textbf{(b)} Based on Widom's expansion for the cavity distribution function\,\cite{cavitym3}, it is expected that the short range cavity logarithm can be approximated by an even polynomial, \emph{i.e.} $\ln{[y(r)]}=\sum_{i=0}^{l}y_{2i}r^{2i}$. Least square fitting confirmed that this expansion is very accurate and led to the determination of the coefficients $y_{2i}$. Notice that $\ln{[y(r)]}$ is a monotonic function in the short range. \textbf{(c)} Given that $-\ln{[g(r)]}=-\ln{[y(r)]}+\beta{u}(r)$, it is evident that the improved biasing component becomes $\beta\phi(r)=\ln{[y(r)]}$ or equivalently
\begin{equation}
\beta\phi(x)=\sum_{i=0}^{l}y_{2i}x^{2i}\,.\label{biasingfullequation}
\end{equation}
where $l$ is the minimum number of polynomial terms that ensures an excellent fit to the short range $\ln{[y(r)]}$ data. This expression for the biasing component of the tagged pair potential is valid in the entire short range of interest. However, this does not suggest that the windowing technique should be abandoned in the final ultra-long cavity simulations, since a broader type of windowing is still essential in order to confine the tagged particles within the correlation void.

\begin{table*}
	\caption{Results of the long cavity simulation matching procedure for all the $16$ YOCP state points of interest. Matching in the overlapping interval $I_5\bigcap{I}_4=[1.25,1.4]$ leads to the constant $\ln{C_4}$, matching in the overlapping interval $I_4\bigcap{I}_3=[0.8,1.0]$ leads to the constant $\ln{C_3}$, matching in the overlapping interval $I_3\bigcap{I}_2=[0.4,0.6]$ leads to the constant $\ln{C_2}$ and matching in the overlapping interval $I_2\bigcap{I}_1=[0.2,0.4]$ leads to the constant $\ln{C_1}$. In each matching stage, the a priori knowledge of the logarithm of the cavity distribution function in the outer interval allows for its determination in the inner interval, starting from the intermediate and long range region $I_5$ where $\ln{[y(r)]}$ is known from the Ornstein-Zernike inversion method. In particular, the proportionality constant is determined by least square fitting $\ln{\left[y_{I_n}(r)/y_{\mathrm{sim}}(r)\right]}=\ln{C_{n-1}}\,\forall{r}\in{I}_{n}\bigcap{I}_{n-1}$ which allows for determination of $y_{I_{n-1}}(r)$ from $\ln{\left[y_{I_{n-1}}(r)/y_{\mathrm{sim}}(r)\right]}=\ln{C_{n-1}}\,\forall{r}\in{I}_{n-1}$. The extremely low mean absolute relative deviations $e_{\ln{C_n}}\,(<0.04\%)$ confirm the theoretical expectation that the ratio $\ln{\left[y(r)/y_{\mathrm{sim}}(r)\right]}$ is well-approximated by a constant.}\label{matchingtable}
	\centering
	\begin{tabular}{cccccccccccc}\hline
$\Gamma_{\mathrm{ISO}}^{\mathrm{OCP}}$ & $\kappa$    & $\Gamma$             & $\ln{C_4}$          & $e_{\ln{C_4}}$      & $\ln{C_3}$          & $e_{\ln{C_3}}$      & $\ln{C_2}$          & $e_{\ln{C_2}}$      & $\ln{C_1}$          & $e_{\ln{C_1}}$             \\ \hline
160\,\,\,                          & \,\,\,1.0\,\,\, &  \,\,\,205.061\,\,\, & \,\,\,235.730\,\,\, & \,\,\,0.012\%\,\,\, & \,\,\,244.758\,\,\, & \,\,\,0.042\%\,\,\, & \,\,\,261.285\,\,\, & \,\,\,0.010\%\,\,\, & \,\,\,261.440\,\,\, & \,\,\,0.006\%\,\,\,          \\
160\,\,\,                          & \,\,\,1.5\,\,\, &  \,\,\,286.437\,\,\, & \,\,\,302.777\,\,\, & \,\,\,0.012\%\,\,\, & \,\,\,316.536\,\,\, & \,\,\,0.008\%\,\,\, & \,\,\,325.572\,\,\, & \,\,\,0.006\%\,\,\, & \,\,\,325.385\,\,\, & \,\,\,0.006\%\,\,\,          \\
160\,\,\,	                       & \,\,\,2.0\,\,\, &  \,\,\,435.572\,\,\, & \,\,\,441.541\,\,\, & \,\,\,0.003\%\,\,\, & \,\,\,455.055\,\,\, & \,\,\,0.005\%\,\,\, & \,\,\,463.919\,\,\, & \,\,\,0.005\%\,\,\, & \,\,\,463.091\,\,\, & \,\,\,0.005\%\,\,\,          \\
160\,\,\,	                       & \,\,\,2.5\,\,\, &  \,\,\,708.517\,\,\, & \,\,\,715.668\,\,\, & \,\,\,0.002\%\,\,\, & \,\,\,728.484\,\,\, & \,\,\,0.005\%\,\,\, & \,\,\,729.232\,\,\, & \,\,\,0.003\%\,\,\, & \,\,\,728.703\,\,\, & \,\,\,0.002\%\,\,\,          \\
120\,\,\,	                       & \,\,\,1.0\,\,\, &  \,\,\,153.796\,\,\, & \,\,\,170.868\,\,\, & \,\,\,0.009\%\,\,\, & \,\,\,183.632\,\,\, & \,\,\,0.012\%\,\,\, & \,\,\,192.335\,\,\, & \,\,\,0.012\%\,\,\, & \,\,\,191.752\,\,\, & \,\,\,0.010\%\,\,\,          \\
120\,\,\,	                       & \,\,\,1.5\,\,\, &  \,\,\,215.930\,\,\, & \,\,\,227.321\,\,\, & \,\,\,0.008\%\,\,\, & \,\,\,233.451\,\,\, & \,\,\,0.009\%\,\,\, & \,\,\,241.817\,\,\, & \,\,\,0.010\%\,\,\, & \,\,\,240.948\,\,\, & \,\,\,0.008\%\,\,\,          \\
120\,\,\,	                       & \,\,\,2.0\,\,\, &  \,\,\,328.816\,\,\, & \,\,\,302.777\,\,\, & \,\,\,0.012\%\,\,\, & \,\,\,316.536\,\,\, & \,\,\,0.008\%\,\,\, & \,\,\,325.572\,\,\, & \,\,\,0.006\%\,\,\, & \,\,\,325.385\,\,\, & \,\,\,0.006\%\,\,\,          \\
120\,\,\,	                       & \,\,\,2.5\,\,\, &  \,\,\,534.722\,\,\, & \,\,\,535.023\,\,\, & \,\,\,0.002\%\,\,\, & \,\,\,539.746\,\,\, & \,\,\,0.003\%\,\,\, & \,\,\,546.893\,\,\, & \,\,\,0.005\%\,\,\, & \,\,\,545.983\,\,\, & \,\,\,0.004\%\,\,\,          \\
80\,\,\,	                       & \,\,\,1.0\,\,\, &  \,\,\,102.531\,\,\, & \,\,\,111.524\,\,\, & \,\,\,0.012\%\,\,\, & \,\,\,115.741\,\,\, & \,\,\,0.018\%\,\,\, & \,\,\,122.975\,\,\, & \,\,\,0.019\%\,\,\, & \,\,\,121.844\,\,\, & \,\,\,0.017\%\,\,\,          \\
80\,\,\,	                       & \,\,\,1.5\,\,\, &  \,\,\,144.330\,\,\, & \,\,\,146.383\,\,\, & \,\,\,0.009\%\,\,\, & \,\,\,149.329\,\,\, & \,\,\,0.015\%\,\,\, & \,\,\,156.405\,\,\, & \,\,\,0.016\%\,\,\, & \,\,\,155.272\,\,\, & \,\,\,0.012\%\,\,\,          \\
80\,\,\,	                       & \,\,\,2.0\,\,\, &  \,\,\,219.972\,\,\, & \,\,\,224.016\,\,\, & \,\,\,0.007\%\,\,\, & \,\,\,227.331\,\,\, & \,\,\,0.009\%\,\,\, & \,\,\,226.880\,\,\, & \,\,\,0.009\%\,\,\, & \,\,\,225.569\,\,\, & \,\,\,0.010\%\,\,\,          \\
80\,\,\,	                       & \,\,\,2.5\,\,\, &  \,\,\,357.136\,\,\, & \,\,\,357.990\,\,\, & \,\,\,0.004\%\,\,\, & \,\,\,361.208\,\,\, & \,\,\,0.006\%\,\,\, & \,\,\,360.429\,\,\, & \,\,\,0.007\%\,\,\, & \,\,\,359.221\,\,\, & \,\,\,0.005\%\,\,\,          \\
40\,\,\,	                       & \,\,\,1.0\,\,\, &  \,\,\,51.265\,\,\,  & \,\,\,53.215\,\,\,  & \,\,\,0.013\%\,\,\, & \,\,\,54.812\,\,\,  & \,\,\,0.023\%\,\,\, & \,\,\,53.619\,\,\,  & \,\,\,0.029\%\,\,\, & \,\,\,52.113\,\,\,  & \,\,\,0.022\%\,\,\,          \\
40\,\,\,	                       & \,\,\,1.5\,\,\, &  \,\,\,72.537\,\,\,  & \,\,\,71.235\,\,\,  & \,\,\,0.009\%\,\,\, & \,\,\,72.242\,\,\,  & \,\,\,0.019\%\,\,\, & \,\,\,70.953\,\,\,  & \,\,\,0.019\%\,\,\, & \,\,\,69.361\,\,\,  & \,\,\,0.016\%\,\,\,          \\
40\,\,\,	                       & \,\,\,2.0\,\,\, &  \,\,\,110.707\,\,\, & \,\,\,106.862\,\,\, & \,\,\,0.006\%\,\,\, & \,\,\,108.041\,\,\, & \,\,\,0.011\%\,\,\, & \,\,\,106.654\,\,\, & \,\,\,0.029\%\,\,\, & \,\,\,105.105\,\,\, & \,\,\,0.039\%\,\,\,          \\
40\,\,\,                           & \,\,\,2.5\,\,\, &  \,\,\,178.269\,\,\, & \,\,\,173.589\,\,\, & \,\,\,0.003\%\,\,\, & \,\,\,174.002\,\,\, & \,\,\,0.009\%\,\,\, & \,\,\,172.470\,\,\, & \,\,\,0.008\%\,\,\, & \,\,\,170.818\,\,\, & \,\,\,0.006\%\,\,\,          \\ \hline
	\end{tabular}
\end{table*}

\begin{table*}
	\caption{\emph{1st-7th column:} Results for the short range Widom representation of the logarithm of cavity distribution function, \emph{i.e.} $\ln{[y(x)]}=y_0+y_2x^2+y_4x^4+\mathcal{O}[x^6]$, for all the $16$ YOCP state points of interest. The low mean absolute relative deviations $\epsilon_{\ln{y}}<0.53\%$ between the MD extracted $\ln{[y(x)]}$ and the truncated Widom series reveal that the first three terms suffice. Note the monotonic dependence of the $y_0,y_2,y_4$ coefficients on the normalized coupling parameter $\Gamma/\Gamma_{\mathrm{m}}$ and screening parameter $\kappa$. \emph{8th-9th column:} Results of the ultra-long cavity simulation matching procedure for all the $16$ YOCP state points. Matching in the overlapping interval $I_5\bigcap{I}=[1.25,1.4]$ where $I=I_4\bigcup{I}_3\bigcup{I}_2\bigcup{I}_1=[0.0,1.4]$ leads to the constant $\ln{C}$. The extremely low mean absolute relative deviations $e_{\ln{C}}\,(<0.005\%)$ again confirm the theoretical expectation that the ratio $\ln{\left[y(r)/y_{\mathrm{sim}}(r)\right]}$ is well-approximated by a constant.}\label{Widomtable}
	\centering
	\begin{tabular}{ccccccc|cc}\hline
$\Gamma_{\mathrm{ISO}}^{\mathrm{OCP}}$ & $\kappa$          & $\Gamma$               & $y_0$               & $y_2$                & $y_4$              & $\epsilon_{\ln{y}}$  & $\ln{C}$              &  $\epsilon_{\ln{C}}$                      \\ \hline
$160$\,\,\,                            & \,\,\,$1.0$\,\,\, &  \,\,\,$205.061$\,\,\, & \,\,\,$76.78$\,\,\, & \,\,\,$-29.49$\,\,\, & \,\,\,$4.06$\,\,\, & \,\,\,$0.12\%$\,\,\, & \,\,\,$199.864$\,\,\, &	\,\,\,$0.002\%$\,\,\,                    \\
$160$\,\,\,                            & \,\,\,$1.5$\,\,\, &  \,\,\,$286.437$\,\,\, & \,\,\,$59.00$\,\,\, & \,\,\,$-25.76$\,\,\, & \,\,\,$3.89$\,\,\, & \,\,\,$0.20\%$\,\,\, & \,\,\,$281.238$\,\,\, &	\,\,\,$0.001\%$\,\,\,                    \\
$160$\,\,\,	                           & \,\,\,$2.0$\,\,\, &  \,\,\,$435.572$\,\,\, & \,\,\,$47.68$\,\,\, & \,\,\,$-22.84$\,\,\, & \,\,\,$3.71$\,\,\, & \,\,\,$0.29\%$\,\,\, & \,\,\,$430.361$\,\,\, &	\,\,\,$0.001\%$\,\,\,                    \\
$160$\,\,\,	                           & \,\,\,$2.5$\,\,\, &  \,\,\,$708.517$\,\,\, & \,\,\,$40.18$\,\,\, & \,\,\,$-20.79$\,\,\, & \,\,\,$3.64$\,\,\, & \,\,\,$0.44\%$\,\,\, & \,\,\,$703.303$\,\,\, &	\,\,\,$0.001\%$\,\,\,                    \\
$120$\,\,\,	                           & \,\,\,$1.0$\,\,\, &  \,\,\,$153.796$\,\,\, & \,\,\,$57.95$\,\,\, & \,\,\,$-22.20$\,\,\, & \,\,\,$3.05$\,\,\, & \,\,\,$0.12\%$\,\,\, & \,\,\,$148.600$\,\,\, &	\,\,\,$0.002\%$\,\,\,                    \\
$120$\,\,\,	                           & \,\,\,$1.5$\,\,\, &  \,\,\,$215.930$\,\,\, & \,\,\,$44.88$\,\,\, & \,\,\,$-19.46$\,\,\, & \,\,\,$2.89$\,\,\, & \,\,\,$0.18\%$\,\,\, & \,\,\,$210.729$\,\,\, &	\,\,\,$0.001\%$\,\,\,                    \\
$120$\,\,\,	                           & \,\,\,$2.0$\,\,\, &  \,\,\,$328.816$\,\,\, & \,\,\,$36.54$\,\,\, & \,\,\,$-17.44$\,\,\, & \,\,\,$2.83$\,\,\, & \,\,\,$0.30\%$\,\,\, & \,\,\,$323.611$\,\,\, &	\,\,\,$0.001\%$\,\,\,                    \\
$120$\,\,\,	                           & \,\,\,$2.5$\,\,\, &  \,\,\,$534.722$\,\,\, & \,\,\,$31.01$\,\,\, & \,\,\,$-16.01$\,\,\, & \,\,\,$2.81$\,\,\, & \,\,\,$0.45\%$\,\,\, & \,\,\,$529.507$\,\,\, &	\,\,\,$0.001\%$\,\,\,                    \\
$80$\,\,\,	                           & \,\,\,$1.0$\,\,\, &  \,\,\,$102.531$\,\,\, & \,\,\,$38.97$\,\,\, & \,\,\,$-14.83$\,\,\, & \,\,\,$2.01$\,\,\, & \,\,\,$0.12\%$\,\,\, & \,\,\,$97.349$\,\,\,  &	\,\,\,$0.003\%$\,\,\,                    \\
$80$\,\,\,	                           & \,\,\,$1.5$\,\,\, &  \,\,\,$144.330$\,\,\, & \,\,\,$30.51$\,\,\, & \,\,\,$-13.24$\,\,\, & \,\,\,$1.99$\,\,\, & \,\,\,$0.21\%$\,\,\, & \,\,\,$139.137$\,\,\, &	\,\,\,$0.002\%$\,\,\,                    \\
$80$\,\,\,	                           & \,\,\,$2.0$\,\,\, &  \,\,\,$219.972$\,\,\, & \,\,\,$25.07$\,\,\, & \,\,\,$-11.94$\,\,\, & \,\,\,$1.94$\,\,\, & \,\,\,$0.32\%$\,\,\, & \,\,\,$214.771$\,\,\, &	\,\,\,$0.001\%$\,\,\,                    \\
$80$\,\,\,	                           & \,\,\,$2.5$\,\,\, &  \,\,\,$357.136$\,\,\, & \,\,\,$21.46$\,\,\, & \,\,\,$-11.02$\,\,\, & \,\,\,$1.92$\,\,\, & \,\,\,$0.43\%$\,\,\, & \,\,\,$351.926$\,\,\, &	\,\,\,$0.003\%$\,\,\,                    \\
$40$\,\,\,	                           & \,\,\,$1.0$\,\,\, &  \,\,\,$51.265$\,\,\,  & \,\,\,$19.84$\,\,\, & \,\,\,$-7.51$\,\,\,  & \,\,\,$1.01$\,\,\, & \,\,\,$0.13\%$\,\,\, & \,\,\,$46.100$\,\,\,  &	\,\,\,$0.005\%$\,\,\,                    \\
$40$\,\,\,	                           & \,\,\,$1.5$\,\,\, &  \,\,\,$72.537$\,\,\,  & \,\,\,$15.82$\,\,\, & \,\,\,$-6.82$\,\,\,  & \,\,\,$1.01$\,\,\, & \,\,\,$0.20\%$\,\,\, & \,\,\,$67.364$\,\,\,  &	\,\,\,$0.003\%$\,\,\,                    \\
$40$\,\,\,	                           & \,\,\,$2.0$\,\,\, &  \,\,\,$110.707$\,\,\, & \,\,\,$13.37$\,\,\, & \,\,\,$-6.52$\,\,\,  & \,\,\,$1.11$\,\,\, & \,\,\,$0.53\%$\,\,\, & \,\,\,$105.519$\,\,\, &	\,\,\,$0.002\%$\,\,\,                    \\
$40$\,\,\,                             & \,\,\,$2.5$\,\,\, &  \,\,\,$178.269$\,\,\, & \,\,\,$11.52$\,\,\, & \,\,\,$-5.88$\,\,\,  & \,\,\,$1.01$\,\,\, & \,\,\,$0.42\%$\,\,\, & \,\,\,$173.073$\,\,\, &	\,\,\,$0.001\%$\,\,\,                    \\ \hline
	\end{tabular}
\end{table*}

\subsection{Implementation of the simulation method}\label{SubCavityImplem}

\noindent The canonical molecular dynamics simulations for the extraction of the cavity distribution functions in the short range were carried out on graphics cards with the RUMD open-source software\,\cite{RUMDref1}. A small number of tests was performed on processors with the LAMMPS package\,\cite{RUMDref2}.

The NVT MD simulations truncate the Yukawa pair potential at $r_{\mathrm{cut}}=8d$ by applying the shifted-force cutoff method\,\cite{RUMDref3}. The numerical time-step employed for the propagation of equations of motion is $\Delta{t}/\tau=2.5\times10^{-3}$ where $\tau=n^{-1/3}\sqrt{m/(k_{\mathrm{B}}T)}$ is the time required for a free particle that is streaming with the thermal velocity to traverse a mean-cubic inter-particle distance. The MD equilibration time is $2^{19}\Delta{t}$ and the configuration saving period is $2^{7}\Delta{t}$, whereas the statistics duration and thus also the number of statistically independent configurations varied strongly, depending on the simulation type. The total particle number is $N=1000$ (998 type A particles and 2 type B particles) leading to $L\simeq16d$ for the cubic simulation box length. The bin width size employed in the histogram method is $\Delta{r}/d=0.002$. Four overlapping windows are employed for the tagged particles, namely $I_1=[0.0,0.4]$, $I_2=[0.2,0.6]$, $I_3=[0.4,1.0]$ and $I_4=[0.8,1.4]$, while we have $I_5=[1.25,\infty)$ for the results that are available from the OZ inversion method. Note that the windowing intervals $I_n$ are always reported in normalized units $x=r/d$.

It is important to justify the relatively low simulated particle number ($N=1000$). In the cavity simulations, the radial distribution function $g_{\mathrm{sim}}^{12}(r)$ of the two tagged particles is extracted. Consequently, a single correlation event is recorded per statistically independent configuration rather than the $N(N-1)/2$ correlation events per configuration that are recorded during the extraction of the radial distribution function $g(r)$ of the targeted system. Since the statistical sample cannot be increased by simulating a larger number of particles, $N$ should be kept as low as possible in order to reduce the computational cost. Nevertheless, in case of very low particle numbers, not only finite size effects would begin to affect the short range static correlations, but the potential of mean force of the simulated system would also start to diverge from the potential of mean force of the targeted system owing to the presence of the two tagged particles. The conservative choice of $N=1000$ ensures that such errors are negligible. The $N-$dependence of the cavity distribution functions has been thoroughly studied at different YOCP state points and will be reported in the section dedicated to uncertainty propagation.

\begin{figure}
	\centering
	\includegraphics[width=3.40in]{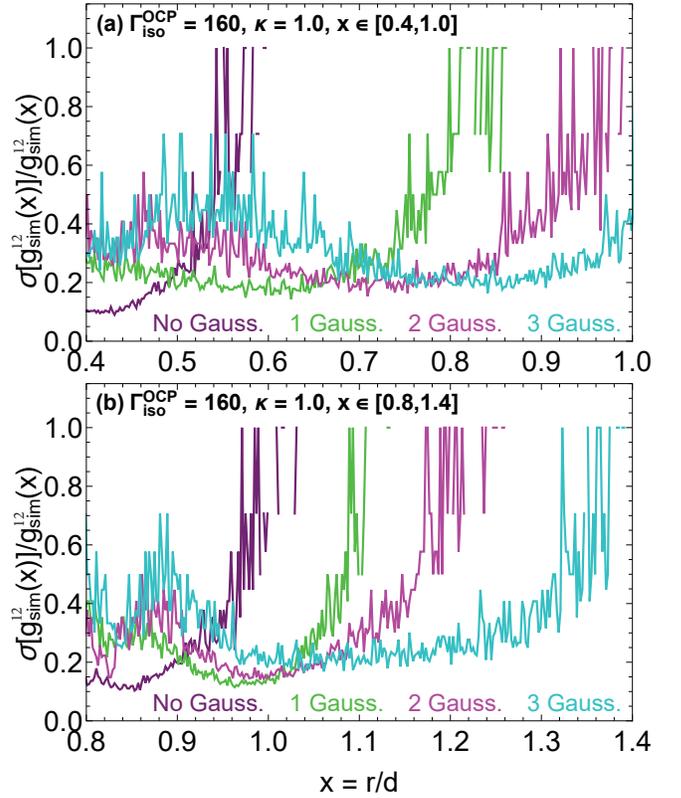}
	\caption{The relative standard errors $\sigma[g_{\mathrm{sim}}^{12}(r)]/g_{\mathrm{sim}}^{12}(r)$ in the extraction of the radial distribution function of the two tagged particles $g_{\mathrm{sim}}^{12}(r)$ from \emph{short cavity simulations} for different biasing components of the tagged pair potential. Results for the YOCP state point $\Gamma_{\mathrm{ISO}}^{\mathrm{OCP}}=160,\,\kappa=1.0$ in the windowing intervals (a) $I_3=[0.4,1.0]$, (b) $I_4=[0.8,1.4]$. In both cases, as the iterative procedure progresses and the biasing component $\beta\phi_n(x)$ is refined by the addition of successive Gaussian functions, $\sigma[g_{\mathrm{sim}}^{12}(r)]/g_{\mathrm{sim}}^{12}(r)$ gradually becomes more uniform in the respective windowing interval. Given the small $M_1=2^{12}$ sample size, the $\sigma[g_{\mathrm{sim}}^{12}(r)]/g_{\mathrm{sim}}^{12}(r)\sim0.3$ level is too high for reliable computation of short range bridge functions, but will be greatly reduced in the long \& ultra-long cavity simulations.}\label{fig:gaussians}
\end{figure}

In the \emph{short cavity simulations}, that are dedicated to the determination of the unknown coefficients of the windowing component of the tagged pair potential and of the Gaussian representation of the biasing component of the tagged pair potential, see Eqs.(\ref{windowingequation},\ref{biasingequation}), the statistics duration is $2^{19}\Delta{t}$ leading to $M_1=2^{12}$ for the statistically independent configuration number. For an arbitrary window $I_{n}=[b_n,c_n]$, the coefficients of the windowing component were found to be $a_1(\Gamma,\kappa)=(25-360)$ depending on the state point, $a_2=20$ regardless of the state point or the confinement interval and $a_{3n}=b_n-1$, $a_{4n}=c_n+1$. Up to three truncations of the iterative procedure proved to be successful in obtaining a Gaussian series representation of the biasing component that allows for the relatively uniform sampling of any windowing interval. Figure \ref{fig:gaussians} demonstrates how the tagged pair statistics become more uniform as the iterative procedure progresses.

\begin{figure*}
	\centering
	\includegraphics[width=7.10in]{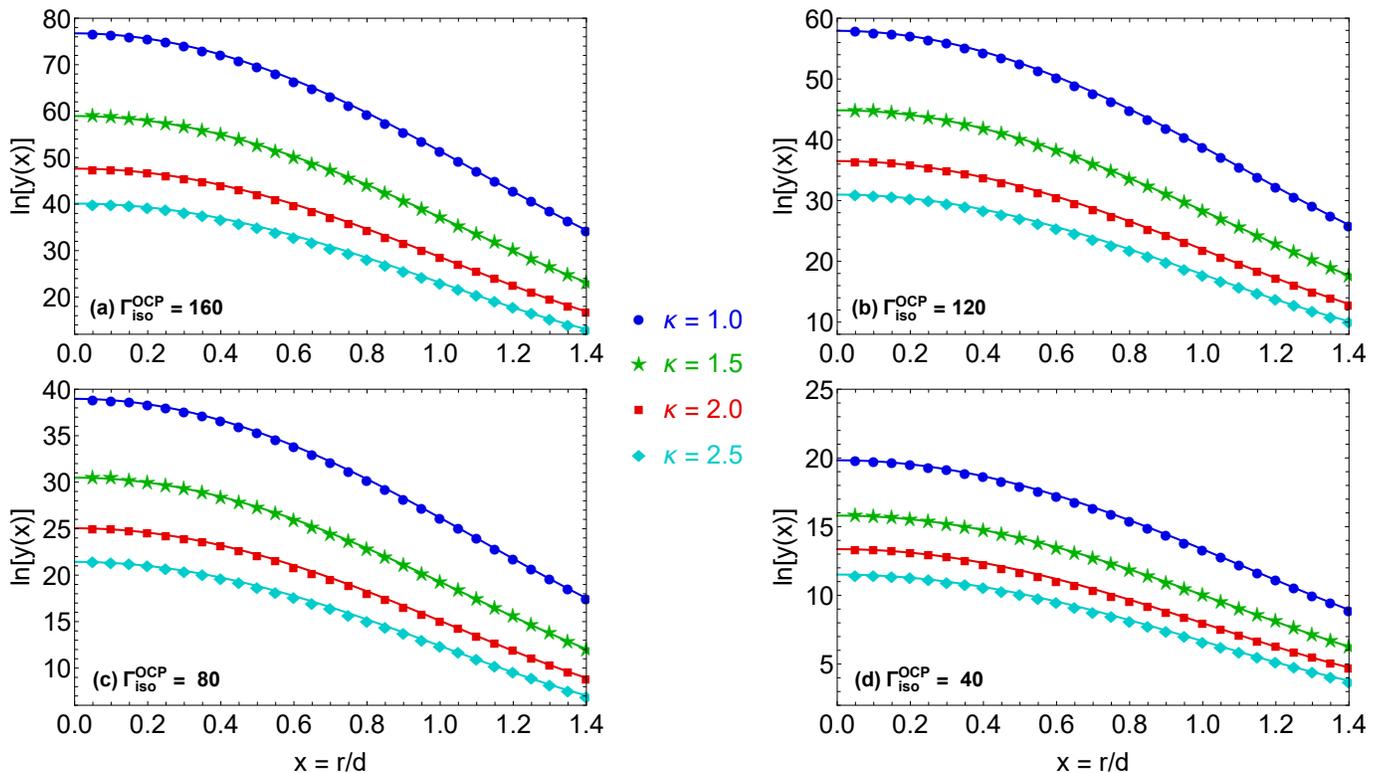}
	\caption{The logarithm of the YOCP cavity distribution function within the correlation void as computed with static correlation input from the \emph{long cavity MD simulations}. Results along the (a) $\Gamma_{\mathrm{ISO}}^{\mathrm{OCP}}=160$ isomorph ($\kappa=1.0,1.5,2.0,2.5$), (b) $\Gamma_{\mathrm{ISO}}^{\mathrm{OCP}}=120$ isomorph ($\kappa=1.0,1.5,2.0,2.5$), (c) $\Gamma_{\mathrm{ISO}}^{\mathrm{OCP}}=80$ isomorph ($\kappa=1.0,1.5,2.0,2.5$), (d) $\Gamma_{\mathrm{ISO}}^{\mathrm{OCP}}=40$ isomorph ($\kappa=1.0,1.5,2.0,2.5$). The $\ln{[y(r/d)]}$ curves as obtained from the MD simulations (discrete symbols) and the Widom expansion with the least square fitted coefficients reported in Table \ref{Widomtable} (solid lines). It is evident that the first three terms of the Widom expansion provide a very accurate representation of $\ln{[y(r/d)]}$ within the correlation void. The MD results, that are extracted every $0.002d$ with the histogram method, have been down-sampled for the purposes of illustration.}\label{fig:cavity-long}
\end{figure*}

In the \emph{long cavity simulations}, that are dedicated to the determination of the unknown coefficients of the Widom polynomial representation of the biasing potential component, see Eq.(\ref{biasingfullequation}), the statistics duration is $2^{27}\Delta{t}$ leading to $M_2=2^{20}$ for the number of statistically independent configurations. The matching procedure, that was followed in the overlapping extent of consecutive confinement windows starting from $I_5\bigcap{I}_4$ and proceeding up to $I_2\bigcap{I}_1$, led to very accurate proportionality constants $C_i$ that are documented in Table \ref{matchingtable}. Note that, due to the extremely large cavity values that emerge in the correlation void, it is preferable to work with natural logarithms. This transforms Eq.(\ref{cavityequation}) into $\ln{\left[y(r)/y_{\mathrm{sim}}(r)\right]}=\ln{C}$ with $y_{\mathrm{sim}}(r)=g_{\mathrm{sim}}^{12}(r)\exp{[\beta{\psi}(r)]}$. Least squares fitting for the cavity logarithm  $\ln{[y(r)]}$ revealed that the first three terms of the Widom expansion suffice for a very accurate representation in the short range. In particular, the absolute relative errors were always less than $0.53\%$. The polynomial coefficients $y_0,y_2,y_4$ are reported in Table \ref{Widomtable}. As illustrated in figure \ref{fig:cavity-long}, the Widom expansion is valid within the entire correlation void and will be ultimately utilized as the biasing component of the tagged pair potential in the final cavity simulations.

\begin{figure}
	\centering
	\includegraphics[width=3.40in]{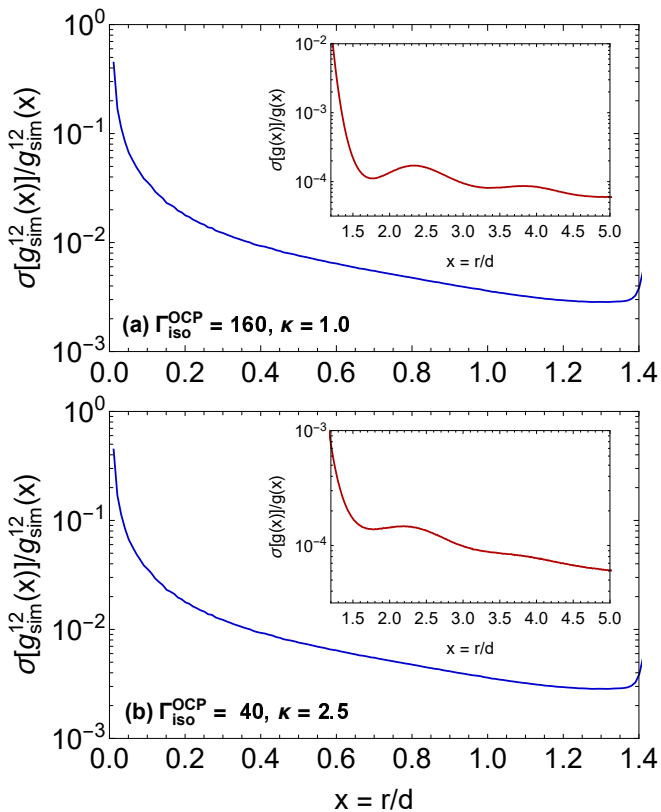}
	\caption{Results for the YOCP state points (a) $\Gamma_{\mathrm{ISO}}^{\mathrm{OCP}}=160$, $\kappa=1.0$ and (b) $\Gamma_{\mathrm{ISO}}^{\mathrm{OCP}}=40$, $\kappa=2.5$. \textbf{Main figure}: Relative standard errors $\sigma[g_{\mathrm{sim}}^{12}(r)]/g_{\mathrm{sim}}^{12}(r)$ in the extraction of the radial distribution function of the tagged particle pair $g_{\mathrm{sim}}^{12}(r)$ from \emph{ultra-long cavity MD simulations} ($N=1000,M_3=2^{25}$) within the short range $r\leq1.4d$.\,\textbf{Inset plot:}\,Relative standard errors $\sigma[g(r)]/g(r)$ in the extraction of the radial distribution function $g(r)$ of the targeted system from \emph{standard NVT MD simulations} ($N=54872,\,M=2^{16}$) in the intermediate range $1.25\leq{r}/d\leq5$\,\cite{accompan}. The error level $\sigma[g_{\mathrm{sim}}^{12}(r)]/g_{\mathrm{sim}}^{12}(r)\sim0.01$ that emerges from the ultra-long cavity simulations is small enough to guarantee a reliable computation of the bridge function in the short range and is comparable to the error level of $\sigma[g(r)]/g(r)$ in the overlapping interval of $1.25\leq{r}/d<1.4$. As the distance decreases, the error level slowly increases up to $r\sim0.2d$, then it begins to steeply increase and finally becomes too high for $r<0.05d$. This suggests that large bridge function uncertainties are localized close to the origin $r=0$. The achieved tagged pair statistics are nearly uniform in the interval $0.2\leq{r}/d\leq1.4$.}\label{fig:notgaussians}
\end{figure}

In the \emph{ultra-long cavity simulations}, that are dedicated to the accurate determination of the cavity distribution function and the bridge function in the short range, the statistics duration is $2^{32}\Delta{t}$ leading to $M_3=2^{25}$ for the number of statistically independent configurations.
The tagged pair potential still features a windowing component that confines the type B particles within the correlation void $I=[0.0,1.4]$. The windowing coefficients were found to be $A_1(\Gamma,\kappa)=(25-360)$ depending on the state point, $A_2=20$, $A_{3}=-0.1$ and $A_{4}=1.5$. The matching procedure, that was followed in the overlapping extent of $I_5\bigcap{I}$, led to very accurate proportionality constants $C$ that are documented in the last two columns of Table \ref{Widomtable}. Figure \ref{fig:notgaussians} demonstrates how the Widom polynomial representation of the biasing potential leads to nearly uniform tagged pair statistics and it also confirms that the chosen size of the final statistical sample suffices for the accurate computation of the short range bridge function in nearly the full extent of the correlation void.

Our version of the cavity simulation method was successfully benchmarked against published results for dense simple liquids. In particular, the short range bridge functions were compared with tabulated MC \& MD simulation data for different state points of Lennard-Jones fluids\,\cite{cavitym2,cavitym6} and tabulated MC simulation data for different state points of inverse power law - $n=12$ - systems\,\cite{cavitym7}. Our MD results always nearly overlapped with the literature results with the exception of minor deviations that lied well within the estimated statistical uncertainties. Unfortunately, such validation exercise was not possible for Yukawa systems, since only MC generated short range screening potentials and not bridge functions are available for the dense YOCP\,\cite{cavitym8}.

\section{Bridge functions determined by the cavity distribution method}\label{CavityBridge}

\subsection{Bridge function computation within the correlation void}\label{SubBridgeVoid}

\begin{figure*}
	\centering
	\includegraphics[width=6.95in]{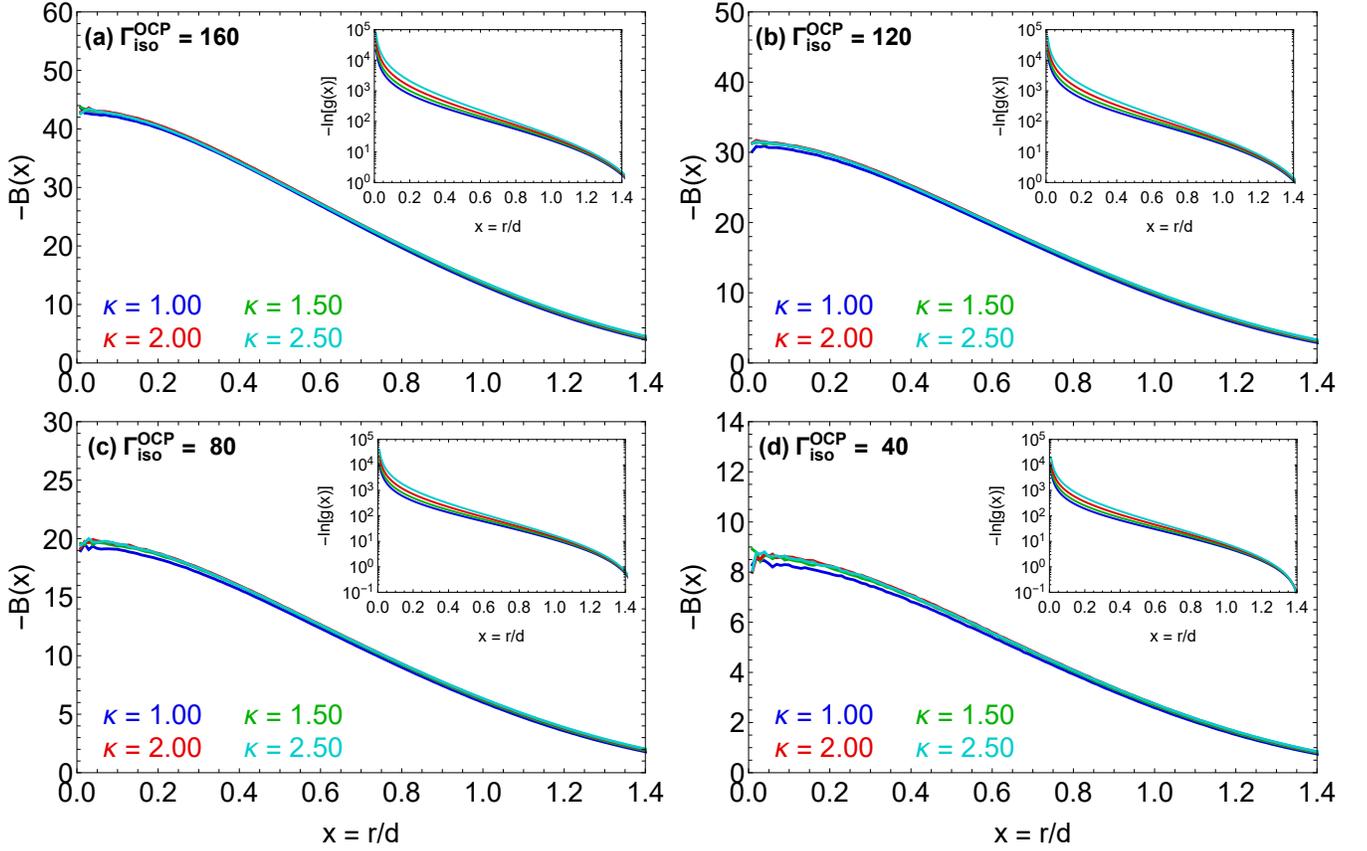}
	\caption{The bridge functions (main) and the potentials of mean force (inset) within the correlation void $r/d\leq1.4$, as computed by applying the cavity distribution method for the YOCP with input from the \emph{ultra-long cavity MD simulations}. Results for: (a) the four members of the $\Gamma_{\mathrm{ISO}}^{\mathrm{OCP}}=160$ isomorph ($\kappa=1.0,1.5,2.0,2.5$); (b) the four members of the $\Gamma_{\mathrm{ISO}}^{\mathrm{OCP}}=120$ isomorph ($\kappa=1.0,1.5,2.0,2.5$); (c) the four members of the $\Gamma_{\mathrm{ISO}}^{\mathrm{OCP}}=80$ isomorph ($\kappa=1.0,1.5,2.0,2.5$); (d) the four members of the $\Gamma_{\mathrm{ISO}}^{\mathrm{OCP}}=40$ isomorph ($\kappa=1.0,1.5,2.0,2.5$).}\label{fig:bridgefunction_lngr}
\end{figure*}

\begin{figure*}
	\centering
	\includegraphics[width=6.95in]{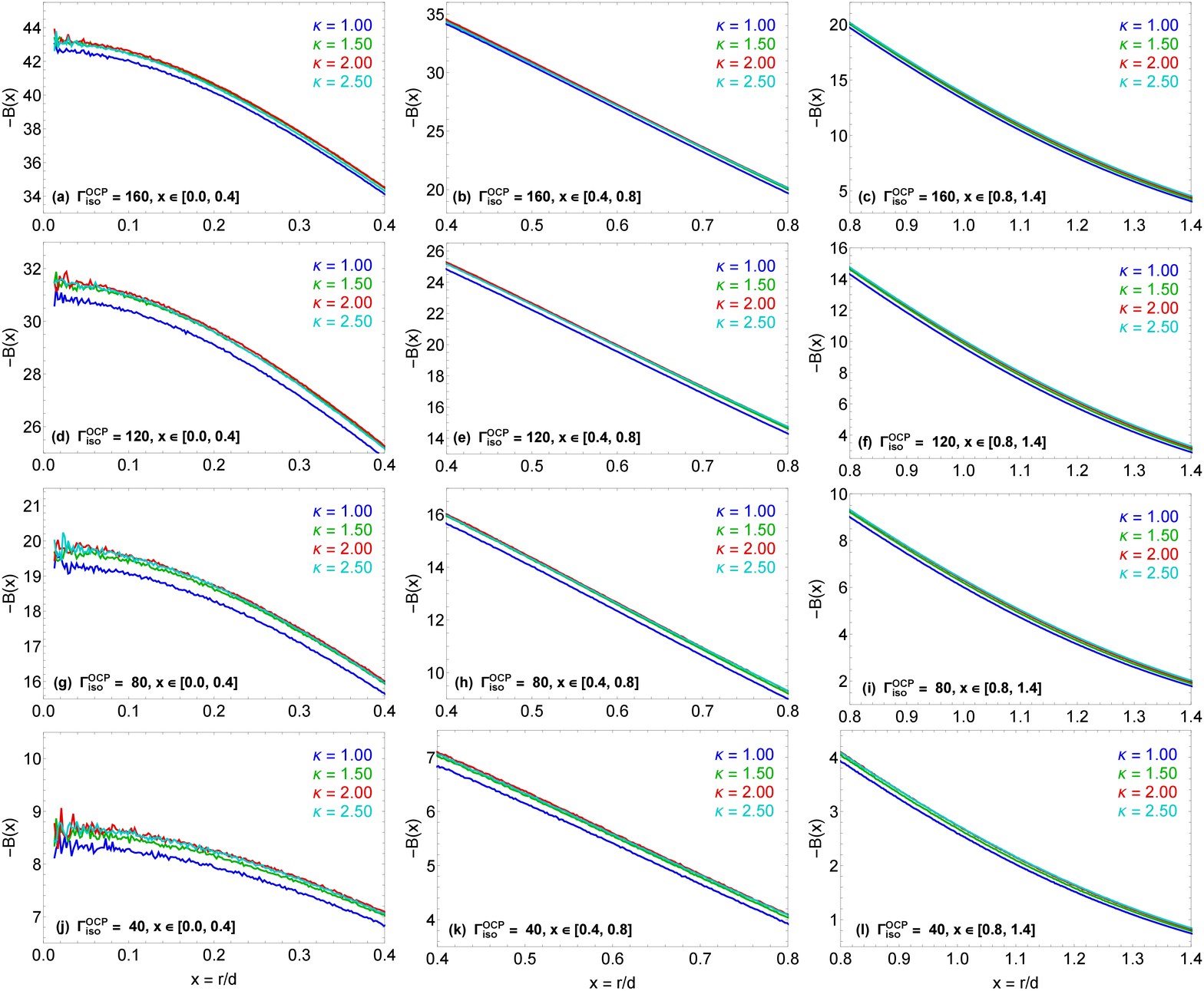}
	\caption{The bridge functions within different sub-intervals of the correlation void $r/d\leq1.4$, as computed by applying the cavity distribution method for the YOCP with input from the \emph{ultra-long cavity MD simulations}. Results for: (a,b,c) the four members of the $\Gamma_{\mathrm{ISO}}^{\mathrm{OCP}}=160$ isomorph ($\kappa=1.0,1.5,2.0,2.5$) in the sub-intervals $[0,0.4d]$, $[0.4d,0.8d]$ and $[0.8d,1.4d]$; (d,e,f) the four members of the $\Gamma_{\mathrm{ISO}}^{\mathrm{OCP}}=120$ isomorph ($\kappa=1.0,1.5,2.0,2.5$) in the sub-intervals $[0,0.4d]$, $[0.4d,0.8d]$ and $[0.8d,1.4d]$; (g,h,i) the four members of the $\Gamma_{\mathrm{ISO}}^{\mathrm{OCP}}=80$ isomorph ($\kappa=1.0,1.5,2.0,2.5$) in the sub-intervals $[0,0.4d]$, $[0.4d,0.8d]$ and $[0.8d,1.4d]$; (j,k,l) the four members of the $\Gamma_{\mathrm{ISO}}^{\mathrm{OCP}}=40$ isomorph ($\kappa=1.0,1.5,2.0,2.5$) in the sub-intervals $[0,0.4d]$, $[0.4d,0.8d]$ and $[0.8d,1.4d]$.}\label{fig:bridgefunction_zoom}
\end{figure*}

\noindent The short range bridge functions $B(r/d)$, as well as the short range potentials of mean force $-\ln{[g(r/d)]}$ as computed from the application of the cavity distribution method with input from the ultra-long MD cavity simulations, are shown in figure \ref{fig:bridgefunction_lngr} for the $4$ isomorphic curves and for the $16$ YOCP state points of interest. The bridge functions are also shown in figure \ref{fig:bridgefunction_zoom} along different sub-intervals of the correlation void.

It is evident that the bridge function is a strongly invariant quantity within the entire correlation void along any YOCP isomorphic curve. This suggests that the isomorph variant short range features of the positive cavity distribution function $\ln{[y(r/d)]}$ (see figure \ref{fig:cavity-long}) and of the negative direct correlation function $c(r/d)$ (see Ref.\cite{accompan}) cancel each other out to a large extent, when these static correlations are added for the computation of the bridge function, see Eq.(\ref{bridgeequation}). The observed degree of bridge function isomorph invariance is high enough that zoom-ins on different sub-intervals are necessary in order to discern the small $B(r/d)$ deviations between the different members of the same isomorph. In addition, it becomes apparent from figure \ref{fig:bridgefunction_zoom} that, regardless of the isomorph, the $\kappa=1.5,\,2.0,\,2.5$ members have nearly overlapping bridge functions whereas the $\kappa=1.0$ member consistently has a slightly displaced bridge function. It is important to notice that the level of invariance is nearly independent of the distance within the entire correlation void, since the relative $B(r)$ deviations between isomorphic state points vary within $2\%-7\%$ for $0<r/d\leq1.4$ (using the $\kappa=1.0$ member as the reference state). As discussed in Ref.\cite{accompan} for the intermediate and long ranges, the small deviations between the bridge functions of isomorphic state points might be comparable to the omnipresent uncertainties in bridge function determination. Therefore, in order to accurately quantify the level of isomorph invariance of the bridge function in the short range, a detailed analysis of all different uncertainty sources is required. This will be pursued in section \ref{UncertaintyPropagation}.

It is worth pointing out that, for a given reduced distance, our methodology leads to statistical errors in the bridge function that are nearly independent of the state point, which implies that the relative errors in the bridge function are largest along the $\Gamma_{\mathrm{ISO}}^{\mathrm{OCP}}=40$ isomorph where the magnitude of the bridge function is the smallest. This expectation is confirmed in the first panel (a,d,g,j) of figure \ref{fig:bridgefunction_zoom}. Furthermore, the large bridge function uncertainties that should arise in the vicinity of the origin $r=0$, as anticipated from the enhanced $\sigma[g_{\mathrm{sim}}^{12}(r)]/g_{\mathrm{sim}}^{12}(r)$ error level recorded close to $r=0$ (see figure \ref{fig:notgaussians}), manifest themselves in the loss of $B(r)$ smoothness and the emergence of spiky saw-toothed $B(r)$ features at $r\lesssim0.05d$.

On the other hand, the potential of mean force is a strongly variant quantity in the entire correlation void along any YOCP isomorphic curve. Notice that the inset plots of figure \ref{fig:bridgefunction_lngr} use logarithmic vertical scales in order to accommodate the five orders of magnitude change in the $\ln{[g(r)]}$ values that takes place from the edge of the correlation void up to the vicinity of the origin, where it ultimately diverges. The $\ln{[g(r/d)]}$ deviations between different members of the same isomorph even exceed one order of magnitude and monotonically increase towards the origin. The above demonstrates that the $g(r/d)$ deviations between different members of the same isomorph are dramatic within the correlation void, in stark contrast to the well-documented isomorph invariance of $g(r/d)$ outside the correlation void\,\cite{isogene7,isogene8}. This was anticipated from the $g(r/d)$ deviations observed between isomorphic state points at the edge of the correlation void, \emph{i.e.} in the interval $1.25\leq{r}/d\leq1.4$ that is reliably probed by the OZ inversion method\,\cite{accompan}.

Finally, for any YOCP state point and within the entire correlation void, the bridge function can be well approximated with a fourth degree polynomial, \emph{i.e.} $B(x)=\sum_{i=0}^{4}b_{i}x^{i}$. The mean absolute relative errors of such fits are $\sim0.5\%$ regardless of the state point. This observation is important in view of attempts to parameterize the dependence of the YOCP bridge function on the $(\Gamma,\kappa)$ state point or attempts to characterize the approximately isomorphic YOCP bridge function exclusively through the reduced excess entropy $s_{\mathrm{ex}}$. This task is beyond the scope of the present work, but will be actively pursued in the future. It is worth noting that such polynomial fits were first employed for the bridge function of hard sphere systems\,\cite{cavitym9}. Note that, in contrast to the logarithm of the cavity distribution function, the short range bridge function cannot be accurately fitted with the Widom even polynomial expansion, which leads to mean absolute relative errors that exceed $\sim7\%$. In view of Eq.(\ref{bridgeequation}), this should be attributed to the small argument expansion of the direct correlation function $c(r)$. In fact, the analytical solution of the soft mean spherical approximation for the direct correlation function within the correlation void contains odd polynomial terms for the YOCP ($\propto{x},x^3$) and contains odd and even polynomial terms for the OCP ($\propto{x}^2,x^3,x^5$)\,\cite{SMSAmina,SMSAminb}.

\subsection{Bridge function extrapolation at the origin}\label{SubBridgeOrigin}

\noindent In the histogram method, the effective bin position lies at the centre of each bin. As a result, the cavity distribution method can be utilized for the computation of the short range bridge function values only at distances larger than $\Delta{r}/2$ ($=0.001d$ in our case), which implies that the value of the bridge function at the origin remains inaccessible. A rigorous small-argument expansion is not available for the bridge function, which suggests that an extrapolation based on the functional behavior of the bridge function at very short distances might lead to erroneous values, especially given the large bridge function uncertainties in the vicinity of $r=0$. Nevertheless, the Widom even polynomial expansion for the cavity distribution function\,\cite{cavitym3} can be utilized for extrapolation purposes. According to Eq.(\ref{bridgeequation}), this would require knowledge of the direct correlation function at the origin. In spite of the absence of a small argument expansion for the direct correlation function, its magnitude at the origin $r=0$ can be computed by taking advantage of the two exact relations of the integral equation theory together with $g(0)=0$ that is valid for origin divergent pair interactions as those realized in YOCP systems.

\begin{table*}
	\caption{The extrapolated value of the bridge function at the origin, $B(0)$, together with all four distinct contributions to its magnitude according to Eq.(\ref{bridgeextrapequation}), for the $16$ YOCP state points of interest. The reduced excess internal energy $u_{\mathrm{ex}}$, the reduced inverse isothermal compressibility $\mu_{\mathrm{T}}$ and the integral residual term $\delta$ are either extracted directly or computed with input from standard canonical MD simulations ($N=54872,\,M=2^{16}$)\,\cite{accompan}. The zero-order term of the Widom expansion $y_0$ is computed by least square fitting input from ultra-long cavity simulations ($N=1000,M_3=2^{25}$) within the short range $r/d\leq1.4$. There are minor deviations $\ll1\%$ between the $y_0$ values that are extrapolated from the ultra-long cavity simulations and those that are extrapolated from the long cavity simulations (see Table \ref{Widomtable}) due to the different statistically independent configurations sampled ($M_3=2^{25}$ versus $M_2=2^{20}$). The contribution of the integral residual term $\delta$ to $B(0)$ is the smallest for all $16$ state points, as expected by the fact that the direct correlation function approaches its asymptotic limit already around $r\simeq2d$ and the fact that the correlation void $g(r)\simeq0$ begins around $r\simeq1.4d$, which imply that the integrand factor $g(x)\left[c(x)+\beta{u}(x)\right]$ is non-zero only within a fraction of the first coordination cell $1.4\lesssim{r}/d\lesssim2$.}\label{ZeroSeptable}
	\centering
	\begin{tabular}{ccccccccc}\hline
$\Gamma_{\mathrm{ISO}}^{\mathrm{OCP}}$ & $\kappa$          & $\Gamma$               & $B(0)$                & $y_0$                & $\mu_{\mathrm{T}}$    & $u_{\mathrm{ex}}$     & $\delta$                               \\ \hline
$160$\,\,\,                            & \,\,\,$1.0$\,\,\, &  \,\,\,$205.061$\,\,\, & \,\,\,$-42.791$\,\,\, & \,\,\,$76.766$\,\,\, & \,\,\,$540.296$\,\,\, & \,\,\,$206.148$\,\,\, & \,\,\,$-7.444$\,\,\,                   \\
$160$\,\,\,                            & \,\,\,$1.5$\,\,\, &  \,\,\,$286.437$\,\,\, & \,\,\,$-43.324$\,\,\, & \,\,\,$58.963$\,\,\, & \,\,\,$284.173$\,\,\, & \,\,\,$88.661$\,\,\,  & \,\,\,$-3.565$\,\,\,                   \\
$160$\,\,\,	                           & \,\,\,$2.0$\,\,\, &  \,\,\,$435.572$\,\,\, & \,\,\,$-43.485$\,\,\, & \,\,\,$47.667$\,\,\, & \,\,\,$193.548$\,\,\, & \,\,\,$48.474$\,\,\,  & \,\,\,$-4.449$\,\,\,                   \\
$160$\,\,\,	                           & \,\,\,$2.5$\,\,\, &  \,\,\,$708.517$\,\,\, & \,\,\,$-43.393$\,\,\, & \,\,\,$40.140$\,\,\, & \,\,\,$150.459$\,\,\, & \,\,\,$30.558$\,\,\,  & \,\,\,$-4.812$\,\,\,                   \\
$120$\,\,\,	                           & \,\,\,$1.0$\,\,\, &  \,\,\,$153.796$\,\,\, & \,\,\,$-30.940$\,\,\, & \,\,\,$57.922$\,\,\, & \,\,\,$405.692$\,\,\, & \,\,\,$155.043$\,\,\, & \,\,\,$-5.745$\,\,\,                   \\
$120$\,\,\,	                           & \,\,\,$1.5$\,\,\, &  \,\,\,$215.930$\,\,\, & \,\,\,$-31.511$\,\,\, & \,\,\,$44.885$\,\,\, & \,\,\,$215.176$\,\,\, & \,\,\,$67.267$\,\,\,  & \,\,\,$-3.246$\,\,\,                   \\
$120$\,\,\,	                           & \,\,\,$2.0$\,\,\, &  \,\,\,$328.816$\,\,\, & \,\,\,$-31.670$\,\,\, & \,\,\,$36.545$\,\,\, & \,\,\,$146.766$\,\,\, & \,\,\,$37.025$\,\,\,  & \,\,\,$-3.501$\,\,\,                   \\
$120$\,\,\,	                           & \,\,\,$2.5$\,\,\, &  \,\,\,$534.722$\,\,\, & \,\,\,$-31.658$\,\,\, & \,\,\,$30.966$\,\,\, & \,\,\,$114.437$\,\,\, & \,\,\,$23.494$\,\,\,  & \,\,\,$-3.824$\,\,\,                   \\
$80$\,\,\,	                           & \,\,\,$1.0$\,\,\, &  \,\,\,$102.531$\,\,\, & \,\,\,$-19.382$\,\,\, & \,\,\,$38.977$\,\,\, & \,\,\,$271.054$\,\,\, & \,\,\,$103.863$\,\,\, & \,\,\,$-3.969$\,\,\,                   \\
$80$\,\,\,	                           & \,\,\,$1.5$\,\,\, &  \,\,\,$144.330$\,\,\, & \,\,\,$-19.785$\,\,\, & \,\,\,$30.483$\,\,\, & \,\,\,$144.489$\,\,\, & \,\,\,$45.467$\,\,\,  & \,\,\,$-2.287$\,\,\,                   \\
$80$\,\,\,	                           & \,\,\,$2.0$\,\,\, &  \,\,\,$219.972$\,\,\, & \,\,\,$-19.948$\,\,\, & \,\,\,$25.046$\,\,\, & \,\,\,$99.007$\,\,\,  & \,\,\,$25.276$\,\,\,  & \,\,\,$-2.462$\,\,\,                   \\
$80$\,\,\,	                           & \,\,\,$2.5$\,\,\, &  \,\,\,$357.136$\,\,\, & \,\,\,$-19.955$\,\,\, & \,\,\,$21.442$\,\,\, & \,\,\,$77.391$\,\,\,  & \,\,\,$16.199$\,\,\,  & \,\,\,$-2.598$\,\,\,                   \\
$40$\,\,\,	                           & \,\,\,$1.0$\,\,\, &  \,\,\,$51.265$\,\,\,  & \,\,\,$-8.3897$\,\,\, & \,\,\,$19.842$\,\,\, & \,\,\,$136.347$\,\,\, & \,\,\,$52.530$\,\,\,  & \,\,\,$-2.055$\,\,\,                   \\
$40$\,\,\,	                           & \,\,\,$1.5$\,\,\, &  \,\,\,$72.537$\,\,\,  & \,\,\,$-8.6413$\,\,\, & \,\,\,$15.798$\,\,\, & \,\,\,$73.552$\,\,\,  & \,\,\,$23.451$\,\,\,  & \,\,\,$-1.211$\,\,\,                   \\
$40$\,\,\,	                           & \,\,\,$2.0$\,\,\, &  \,\,\,$110.707$\,\,\, & \,\,\,$-8.7334$\,\,\, & \,\,\,$13.279$\,\,\, & \,\,\,$50.936$\,\,\,  & \,\,\,$13.320$\,\,\,  & \,\,\,$-1.283$\,\,\,                   \\
$40$\,\,\,                             & \,\,\,$2.5$\,\,\, &  \,\,\,$178.269$\,\,\, & \,\,\,$-8.7523$\,\,\, & \,\,\,$11.534$\,\,\, & \,\,\,$40.021$\,\,\,  & \,\,\,$8.684$\,\,\,   & \,\,\,$-1.366$\,\,\,                  \\ \hline
	\end{tabular}
\end{table*}

The theoretical foundation of the present bridge function extrapolation method strongly resembles that of the so-called zero separation theorems\,\cite{zerosep1,zerosep2,zerosep3}. We set $r=0$ in the Ornstein-Zernike equation, see Eq.(\ref{eq:theory_oz}), and employ $h(0)=-1$. Solving for $c(r=0)$, we acquire
\begin{equation*}
c(0)=-1-n\int\,c(r)h(r)d^3r\,,
\end{equation*}
where we set $\boldsymbol{r}^{\prime}=\boldsymbol{r}$ for the dummy volume integration variable. After substituting for $h(r)=g(r)-1$ and splitting the integral, the statistical relation for the reduced inverse isothermal compressibility $\mu_{\mathrm{T}}=1-n\int\,c(r)d^3r$ emerges. By adding and then subtracting the asymptotic limit of the direct correlation function $-\beta{u}(r)$ in the respective factor of the integrand, the reduced excess internal energy $u_{\mathrm{ex}}=(1/2)n\beta\int\,g(r)u(r)d^3r$ emerges. The remaining integral can be simplified by utilizing spherical coordinates and normalized distance units $x=r/d$. The above leads to
\begin{equation*}
c(0)=-\mu_{\mathrm{T}}+2u_{\mathrm{ex}}-3\int_0^{\infty}x^2g(x)\left[c(x)+\beta{u}(x)\right]dx\,.
\end{equation*}
We proceed with setting $r=0$ in the non-linear closure equation, Eq.(\ref{eq:theory_oz_closure}), and we employ $h(0)=-1$. The singularities of the interaction potential and of the potential of mean force at the origin can be removed by combining the two diverging terms for the finite valued logarithm of the cavity distribution function $\ln{[y(0)]}$ to emerge, whose value at the origin is given by the zero-order term of the Widom expansion $y_0$. Thus, we have
\begin{equation*}
B(0)=y_0+c(0)+1\,.
\end{equation*}
Finally, we combine the above equations and we also set $\delta=3\int_0^{\infty}x^2g(x)\left[c(x)+\beta{u}(x)\right]dx$ to end up with
\begin{equation}
B(0)=y_0-\mu_{\mathrm{T}}+2u_{\mathrm{ex}}-\delta+1\,.\label{bridgeextrapequation}
\end{equation}
In this expression, all four contributions to the $B(0)$ value can be easily determined from available MD simulations: $u_{\mathrm{ex}}$ from the canonical mean of total potential energies, $\mu_{\mathrm{T}}$ from the hypervirial route\cite{accompan,hypervir},\,$\delta$ by combining the Ornstein-Zernike equation with simulation extracted radial distribution functions, $y_0$ by least square fitting the output of the cavity simulations within the correlation void to the first three terms of the Widom series. Note that the more accurate $y_0$ values resulting from the ultra-long cavity simulations should be preferred over the $y_0$ values that resulted from the long cavity simulations and were reported in Table \ref{Widomtable}.

The $B(0)$ values that result from the above extrapolation method are collected in Table \ref{ZeroSeptable}. None of the contributing quantities $(y_0,\,\mu_{\mathrm{T}},\,u_{\mathrm{ex}},\,\delta)$ are isomorph invariant; $y_0$ due to its deep connection with the excess chemical potential\,\cite{zerosep1,zerosep2}, $\mu_{T}$ because its thermodynamic definition involves volume derivatives\,\cite{isogene8}, $u_{\mathrm{ex}}$ owing to its straightforward connection with the interaction potential (see also the Rosenfeld-Tarazona decomposition\,\cite{decompo1,decompo2}), $\delta$ due to the presence of the direct correlation function in the integrand\,\cite{accompan}. The isomorph variance is confirmed in Table \ref{ZeroSeptable}, where large differences can be observed in the above quantities amongst isentropic YOCP state points. In spite of the fact that the $(y_0,\,\mu_{\mathrm{T}},\,u_{\mathrm{ex}},\,\delta)$ contributing quantities are not isomorph invariant, the bridge function at the origin is revealed to be isomorph invariant to a high degree, as expected from the $B(r)$ invariance in the statistically sampled range. In fact, a simple quadratic extrapolation of the bridge function in the smooth interval $0.05<r/d<0.4$ would result in very similar but slightly less accurate values of $B(0)$. Finally, we note that specially designed simulations based on the insertion of test particles can be utilized in order to extract $\ln{[y(0)]}=y_0$ exactly at the origin, thus avoiding the need for any extrapolation\,\cite{zerosep4}. As we shall discuss in section \ref{UncertaintyPropagation}, due to the very high number of statistically independent configurations sampled in the ultra-long cavity simulations, the extrapolated $B(0)$ values are accurate enough so that an additional series of simulations is judged to be redundant.

\section{Propagation of uncertainties}\label{UncertaintyPropagation}

\subsection{Types of errors and sources of uncertainty}\label{SubUncertaintyIntro}

\noindent Paper I featured a detailed analysis of all uncertainties relevant to the computation of bridge functions with the OZ inversion method in the intermediate and long range. Five types of errors were identified; statistical errors due to the finite simulation duration, grid errors due to the finite histogram width, size errors due to the finite number of simulated particles, tail errors due to the finite length of the simulation box and isomorphic errors due to excess entropy mismatches between nominally isomorphic state points\,\cite{accompan}. As emphasized in Paper I, outside the correlation void, the above uncertainties propagate from the radial distribution function to the bridge function mainly through the direct correlation function that is calculated from the OZ equation, see Eq.(\ref{bridgeequationfull}).

For the ultra-accurate canonical MD simulations ($N=54872,\,M=2^{16}$) that were employed for the reliable extraction of the radial distribution function in the range $1.25d\leq{r}\leq30d$, we proceeded in the following manner: (a) the statistical errors were quantified with a block average procedure that divides the total dataset of $M$ uncorrelated configurations into $N_{\mathrm{b}}$ blocks each containing $N_{\mathrm{g}}$ configurations, (b) the grid errors were demonstrated to be negligible compared to the statistical errors due to the narrow $\Delta{r}/d=0.002$ histogram bin widths utilized, (c) the size errors were compensated with the Lebowitz-Percus correction\,\cite{sizeexp1,sizeexp2}, (d) the tail errors were revealed to be negligible by applying different long range extrapolation methods,\,(e)\,the isomorphic errors were also proven to be insignificant, since isomorph tracing with the direct isomorph check and the small step method led to minor uncertainties $\Delta\Gamma/\Gamma<1\%$ in the coupling parameters of isomorphic state points.

The five types of errors listed above are also relevant to the computation of bridge functions with the cavity distribution method in the short range. Combining Eq.(\ref{bridgeequation}) with Eq.(\ref{cavityequation}) and setting $y_{\mathrm{sim}}(r)=\exp{[\beta{\psi}(r)]}g_{\mathrm{sim}}^{12}(r)$ for convenience, we obtain $B(r)=\ln{[y_{\mathrm{sim}}(r)]}+\ln{C}+c(r)+1$. There are naturally uncertainties in the determination of each of the non-trivial terms; uncertainties in $\ln{[y_{\mathrm{sim}}(r)]}$ originate\,from its computation with input from ultra-long cavity MD simulations, uncertainties in $c(r)$ emerge from its computation with input from ultra-accurate MD simulations, uncertainties in $\ln{C}$ stem from its computation with the matching procedure in the overlapping interval of $1.25\leq{r/d}\leq1.4$. Since these sources of uncertainty must be statistically independent, the overall bridge function uncertainties are given by
\begin{equation}
\sigma[B(r)]=\sqrt{\sigma^2[\ln{y_{\mathrm{sim}}(r)}]+\sigma^2[c(r)]+\sigma^2[\ln{C}]}\,,\label{uncertaintyequation}
\end{equation}
where $\sigma[.]$ denotes the standard error of the mean of the quantity or function in brackets.

After applying the methodology described in Paper I, it has been confirmed that the grid errors, the tail errors and the isomorphic errors in the bridge function are negligible also in the correlation void. In addition, it has been revealed that the cavity uncertainties $\sigma[\ln{y_{\mathrm{sim}}(r)}]$ are always dominant compared to the direct correlation uncertainties $\sigma[c(r)]$ and the matching constant uncertainties $\sigma[\ln{C}]$, which indicates that $\sigma[B(r)]\simeq\sigma[\ln{y_{\mathrm{sim}}(r)}]\,,\,\forall{r}$. Thus, in what follows, we will focus on the quantification of the statistical errors and size errors in $\ln{[y_{\mathrm{sim}}(r)]}$, even though the uncertainties in $\ln{C}$ and $c(r)$ will also be considered when calculating the overall bridge function uncertainties. As far as $\sigma[c(r)]$ is concerned, the statistical errors have been quantified with use of $N_{\mathrm{b}}=N_{\mathrm{g}}=256$ blocks and the explicit finite size errors have been compensated for by applying the Lebowitz-Percus correction, see Paper I for details.

\subsection{Statistical errors}\label{SubStatistical}

\noindent In the OZ inversion method, the implementation of the block averaging method for the quantification of the statistical bridge function uncertainties was necessitated by the fact that extremely smooth radial distribution functions are required as input for the computation of meaningful bridge functions (and direct correlation functions). Even for the ultra-accurate NVT MD simulations ($N=54872,\,M=2^{16}$) that feature $N(N-1)/2\simeq1.5\times10^{9}$ pair statistics per configuration spread over a distance of $60d$, the radial distribution functions that were extracted per configuration were not considered as sufficiently smooth for bridge function computation. This smoothness condition ultimately originates from the fact that bridge functions and direct correlation functions lack a probabilistic interpretation.

\begin{figure*}
	\centering
	\includegraphics[width=7.05in]{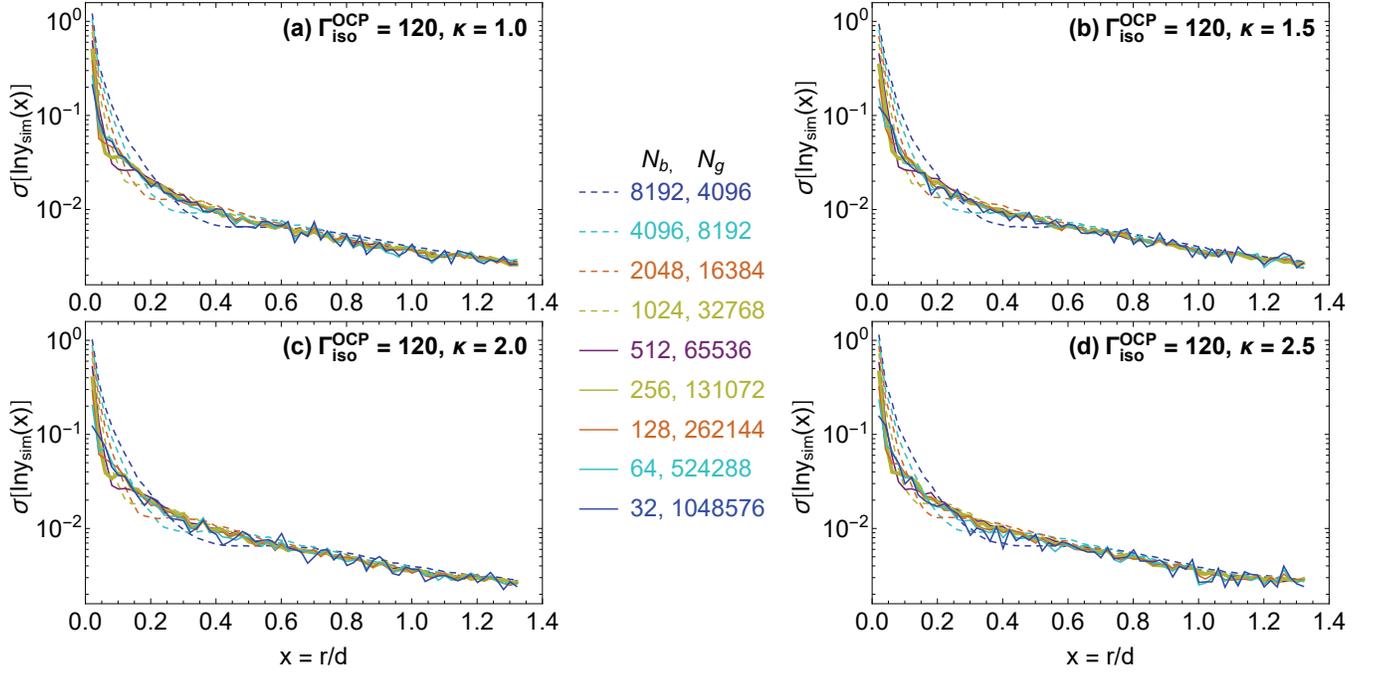}
	\caption{Statistical errors in the extraction of cavity distribution function logarithms $\ln{[y_{\mathrm{sim}}(r)]}$ from \emph{ultra-long cavity MD simulations} along the isomorph $\Gamma_{\mathrm{ISO}}^{\mathrm{OCP}}=120$ ($\kappa=1.0,1.5,2.0,2.5$). Determination for $9$ combinations of the block and sub-block configuration number; $(N_{\mathrm{b}},\,N_{\mathrm{g}})=\{(2^{13},2^{12}),(2^{12},2^{13}),(2^{11},2^{14}),(2^{10},2^{15}),(2^{9},2^{16}),(2^{8},2^{17}),(2^{7},2^{18}),(2^{6},2^{19}),(2^{5},2^{20})\}$. Notice the logarithmic scale of the vertical axis. A strong overlapping of the statistical errors $\sigma[\ln{y_{\mathrm{sim}}(r)}]$ can be discerned for the combinations that feature a large number of sub-block configurations, \emph{i.e.} $(N_{\mathrm{b}},\,N_{\mathrm{g}})=\{(2^{9},2^{16}),(2^{8},2^{17}),(2^{7},2^{18}),(2^{6},2^{19}),(2^{5},2^{20})\}$. Ultimately, the combination $(N_{\mathrm{b}},\,N_{\mathrm{g}})=(2^{5},2^{20})$ was selected for the formal quantification of $\sigma[\ln{y_{\mathrm{sim}}(r)}]$. It is important to observe that the statistical errors are nearly independent of the isomorphic state point.}\label{fig:cavity_statistical_error}
\end{figure*}

In the cavity distribution method, the logarithm of the cavity distribution function has a probabilistic interpretation\footnotemark\footnotetext{Strictly speaking, $\ln{[y_{\mathrm{sim}}(r)]}$ is the logarithm of the cavity distribution function of the simulated system and $\ln{[y(r)]}$ is the logarithm of the cavity distribution function. These quantities are connected with the additive constant, $\ln{[y(r)]}=\ln{[y_{\mathrm{sim}}(r)]}+\ln{C}$, that is determined by the matching procedure, see sections \ref{CavityMD} and \ref{SubUncertaintyIntro} for a detailed discussion. Nonetheless, for the sake of brevity and only in this section, we will refer to $\ln{[y_{\mathrm{sim}}(r)]}$ as the logarithm of the cavity distribution function or cavity logarithm.}.\,However, since the useful statistics are exclusively generated by the tagged particle pair, a unique correlation event at a single distance is sampled per configuration. As a result, meaningful logarithms of the cavity distribution function can only be constructed by combining the statistics of multiple configurations. Hence, a block averaging method should also be implemented in order to quantify the statistical cavity logarithm uncertainties within the short range. In the following, we shall denote MD simulation averages by $\langle...\rangle_{K}$ where $K$ is the number of statistically independent configurations.

The following block averaging procedure was used in order to estimate the statistical uncertainties $\sigma[\ln{y_{\mathrm{sim}}(r)}]$ arising in ultra-long cavity simulations ($N=1000,M_3=2^{25}$). The total dataset of $M_3$ uncorrelated configurations was divided into $N_{\mathrm{b}}$ blocks each containing $N_{\mathrm{g}}$ configurations ($M_3=N_{\mathrm{b}}\times{N}_{\mathrm{g}}$). For each block, a block-averaged cavity logarithm was calculated at all sampled distances, \emph{i.e.} $\langle\ln{[y_{\mathrm{sim},i}(r)]}\rangle_{N_{\mathrm{g}}}=(1/N_{\mathrm{g}})\sum_{j=1}^{N_{\mathrm{g}}}\ln{[y_{\mathrm{sim}}(r,j\Delta{t})]}$. By introducing the $\ln{[y_{\mathrm{sim}}(r)]}=\langle\ln{[y_{\mathrm{sim}}(r)]}\rangle_{M_3}$ notation for the mean cavity logarithm, the standard deviation of the mean was straightforwardly computed by $\sigma^2[\ln{y_{\mathrm{sim}}(r)}]=1/[{N_{\mathrm{b}}(N_{\mathrm{b}}-1)}]\sum_{i=1}^{N_{\mathrm{b}}}\left\{\langle\ln{[y_{\mathrm{sim},i}(r)]}\rangle_{N_{\mathrm{g}}}-\ln{[y_{\mathrm{sim}}(r)]}\right\}^2$.

The $(N_{\mathrm{b}},\,N_{\mathrm{g}})$ combination remains to be determined. The $N_{\mathrm{g}}$ value should be as large as possible to guarantee that $\langle\ln{[y_{\mathrm{sim},i}(r)]}\rangle_{N_{\mathrm{g}}}$ is a sufficiently smooth function in the correlation void (except perhaps in the vicinity of the origin where even the overall statistics are rather poor). The $N_{\mathrm{b}}$ value should be large enough for the sample size of $\langle\ln{[y_{\mathrm{sim},i}(r)]}\rangle_{N_{\mathrm{g}}}$ to be sufficiently large, otherwise the estimate of $\sigma[\ln{y_{\mathrm{sim}}(r)}]$ would not be reliable. Given the single useful statistic per configuration, it is evident that combinations that satisfy $N_{\mathrm{g}}\gg{N}_{\mathrm{g}}$ are preferable.

The effect of different $(N_{\mathrm{b}},N_{\mathrm{g}})$ combinations on the statistical cavity logarithm uncertainties was investigated for all the $16$ YOCP state points of interest. Regardless of the thermodynamic state point, the statistical errors were observed to fluctuate for combinations that feature low values of $N_{\mathrm{g}}(\lesssim50000)$, but they remained approximately constant for those combinations with $N_{\mathrm{g}}\gtrsim50000$, see the logarithmic plots of figure \ref{fig:cavity_statistical_error} for characteristic examples. Ultimately, the combination of $(N_{\mathrm{b}},\,N_{\mathrm{g}})=(32,1048576)$ or equivalently $(N_{\mathrm{b}},\,N_{\mathrm{g}})=(2^{5},2^{20})$ was selected for the quantification of the statistical errors in the cavity logarithm $\ln{[y_{\mathrm{sim}}(r)]}$ which dominate the statistical errors in the bridge function.

It is important to point out that the magnitude of the $\sigma[\ln{y_{\mathrm{sim}}(r)}]$ errors was confirmed to be nearly independent of the YOCP state point. Taking into account that $\sigma[B(r)]\simeq\sigma[\ln{y_{\mathrm{sim}}(r)}]$ and that the lowest bridge function magnitudes emerge along the $\Gamma_{\mathrm{ISO}}^{\mathrm{OCP}}=40$ isomorph, it is concluded that the bridge functions along this isomorph are characterized by the largest relative errors, see also section \ref{SubBridgeVoid}.

\begin{figure}
	\centering
	\includegraphics[width=3.40in]{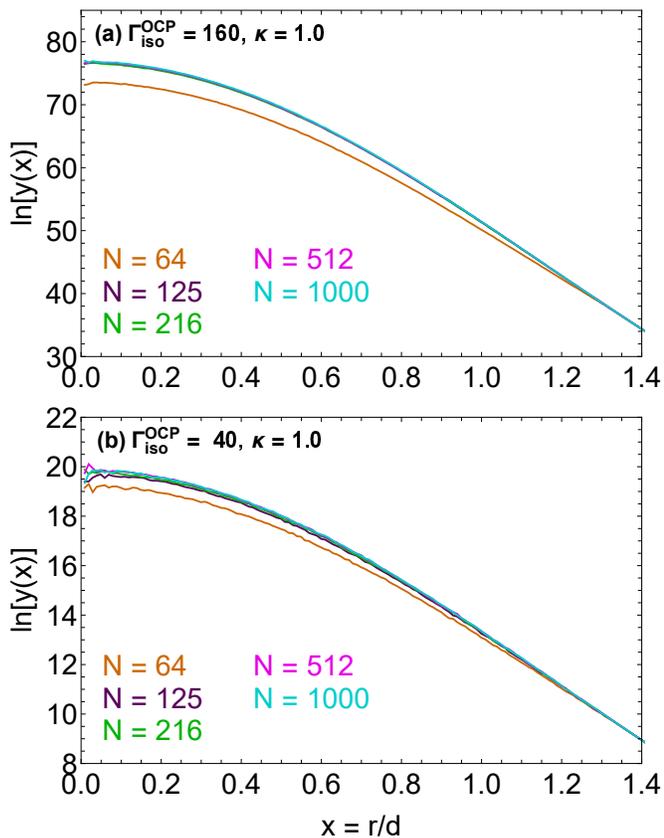}
	\caption{The dependence of the logarithm of the YOCP cavity distribution function $\ln{[y(r)]}$ within the correlation void on the simulated number of particles $N$. Results for five different particle numbers ($N=1000,\,512,\,216,\,125,\,64$) at the YOCP state points (a) $\Gamma=205.061$, $\kappa=1.0$ ($\Gamma_{\mathrm{ISO}}^{\mathrm{OCP}}=160$) and (b) $\Gamma=51.265$, $\kappa=1.0$ ($\Gamma_{\mathrm{ISO}}^{\mathrm{OCP}}=40$). The $\ln{[y(r)]}$ functions were computed with input from \emph{ultra-long cavity MD simulations} ($M_3=2^{25}$) in which the Yukawa pair potential was truncated at $r_{\mathrm{cut}}=L/2$ with the shifted-force cutoff method, where $L$ is the length of the cubic simulation box. Since the number of simulated particles varied between simulations, the box length and, thus, the interaction cutoff radius also varied. In particular, we used $r_{\mathrm{cut}}/d=8.0,\,6.5,\,4.8,\,4.0,\,3.2$. For both YOCP state points, the $\ln{[y(r)]}$ results are nearly indistinguishable within the correlation void when $N=1000,\,512,\,216,\,125$ and observable deviations emerge only when $N=64$. We conclude that finite size effects are negligible for $N\geq125$.}\label{fig:cavity_Ndependence}
\end{figure}

\subsection{Finite size errors}\label{SubFiniteSize}

\noindent In the ultra-accurate standard NVT MD simulations carried out for the computation of the bridge function in the intermediate and the long range, the number of simulated particles was $N=54872$. Nevertheless, it was still revealed that explicit finite size effects managed to distort the asymptotic behavior of bridge functions and they had to be compensated for in order to ensure a proper convergence to zero\,\cite{accompan}. In all three types of cavity NVT MD simulations carried out for the computation of the bridge function in the short range, the number of simulated particles was merely $N=1000$. This is a result of the fact that a single useful correlation event is recorded per statistically independent configuration and it concerns the tagged particle pair. Thus, larger particle numbers would only manifoldly increase the computational cost without improving the statistics. However, as discussed in section \ref{SubCavityImplem}, too low numbers of particles should even make the bridge function within the correlation void susceptible to finite size errors.

Since the direct correlation functions have been computed with input from the ultra-accurate standard NVT MD simulations of $N=54872$ and have already been corrected for finite size errors, it suffices to investigate the $N-$dependence of the cavity distribution function $y(r)$ and more specifically of its logarithm $\ln{[y(r)]}$,\,see\,Eq.(\ref{bridgeequation}). In particular, the cavity distribution function logarithm was computed at different YOCP state points with input from ultra-long cavity MD simulations featuring different particle numbers in an effort to determine the threshold of the $\ln{[y(r)]}$ $N$-dependence. The final $\ln{[y(r)]}$ results revealed that, regardless of the YOCP state point, finite size effects do not emerge at least for $N\geq125$. This conclusion is consistent with an early investigation for dense Lennard-Jones liquids, where the cavity correlation functions in the short range were stated to be nearly identical for $N=133$ and $N=500$\,\cite{cavitym2}. Two characteristic examples are illustrated in figure \ref{fig:cavity_Ndependence}.

Our choice of $N=1000$ is certainly on the conservative side. It is worth pointing out that, with the exception of a few investigations, low numbers of particles are typically considered in the cavity MD or MC simulations reported in the literature. For instance, we have $N=133$\,\cite{cavitym2,cavitym7}, $N=512$\,\cite{cavitym4}, $N=100$\,\cite{cavitym5}, $N=5002$\,\cite{cavitym6}, $N=500$\,\cite{cavitym8}, $N=1000$\,\cite{cavityn1}, $N=4000$\,\cite{cavityn2}.

\section{Bridge functions including uncertainties}\label{UncertaintyBridge}

\begin{figure*}
	\centering
	\includegraphics[width=7.05in]{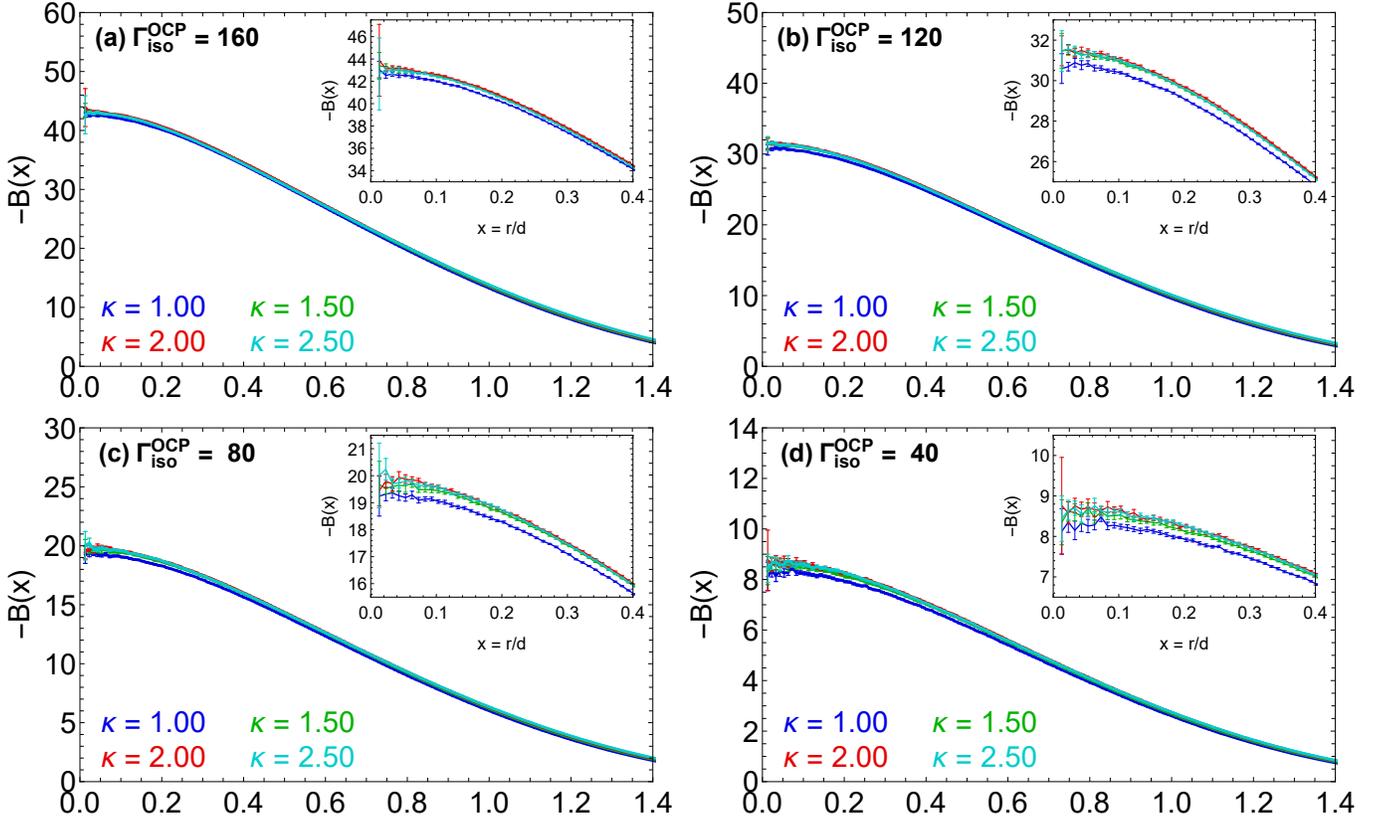}
	\caption{Bridge functions of dense YOCP liquids within the correlation void featuring error bars due to statistical uncertainties ($95\%$ confidence intervals). The short range bridge functions are calculated from $B(r)=\ln{[y(r)]}+c(r)+1$, where the cavity distribution functions $y(r)$ are computed by applying the cavity distribution method with input from \emph{ultra-long canonical MD cavity simulations} and the direct correlation functions $c(r)$ are computed by applying the Ornstein-Zernike inversion method with input from \emph{ultra-accurate canonical MD simulations}. (a,\,b,\,c,\,d) Results for the $4$ members of the $\Gamma_{\mathrm{ISO}}^{\mathrm{OCP}}=160,\,120,\,80,\,40$ isomorphs, respectively. The isomorphic deviations as well as the error bars are very small and can be better discerned in the zero-separation vicinity from the zoomed-in insets.}\label{fig:bridgefunction_errorbars}
\end{figure*}

\noindent The errors bars for the bridge functions within the correlation void are computed in the following manner for all the $16$ YOCP state points of interest. (a) The statistical uncertainties in the direct correlation function $\sigma[c(r)]$ (that was computed with the OZ inversion method) are calculated with the aid of the block averaging method for the combination $(N_{\mathrm{b}},\,N_{\mathrm{g}})=(256,\,256)$, see Paper I for details. (b) The statistical uncertainties in the matching constant logarithm $\sigma[\ln{C}]$ are calculated from the standard deviations of $\ln{C}$ from its mean value for all the histogram bins that belong in the overlapping interval. (c) The statistical uncertainties in the logarithm of the cavity distribution function of the simulated system $\sigma[\ln{y_{\mathrm{sim}}(r)}]$ are calculated with the aid of the block averaging method for the combination $(N_{\mathrm{b}},\,N_{\mathrm{g}})=(32,1048576)$, see section \ref{SubStatistical} for details. (d) The statistical uncertainties in the bridge function $\sigma[B(r)]$ are calculated by combining the contributions of these three uncorrelated sources of uncertainties, see Eq.(\ref{uncertaintyequation}). (e) This standard error of the mean $\sigma[B(r)]$ is utilized for the determination of confidence intervals. In particular, the selected error bars for the statistical uncertainties correspond to $95\%$ confidence intervals.

The short range ($r/d\leq{1.4}$) YOCP bridge functions including error bars due to the statistical uncertainties have been illustrated in figure \ref{fig:bridgefunction_errorbars} for the four isomorphic curves and $16$ state points of interest. By inspecting the bridge functions of the $\kappa=1.5,\,2.0,\,2.5$ members for each isomorph, it is deduced that the statistical uncertainties are comparable to the isomorphic deviations observed. On the other hand, by inspecting the bridge function of the $\kappa=1.0$ member with respect to the bridge functions of the $\kappa=1.5,\,2.0,\,2.5$ members, it is evident that the statistical uncertainties are too minuscule to account even for the small deviations observed. Overall, it is concluded that, similar to the intermediate and the long range, the observed isomorph invariance of the Yukawa bridge functions in the short range is only of approximate nature. In terms of absolute numbers, the degree of isomorph invariance is roughly similar between the intermediate \& long range and the short range. However, in terms of relative numbers, the degree of isomorph invariance is much higher in the short range compared to the intermediate \& long range, as a result of the much higher bridge function magnitude within the correlation void.

\begin{figure*}
	\centering
	\includegraphics[width=6.85in]{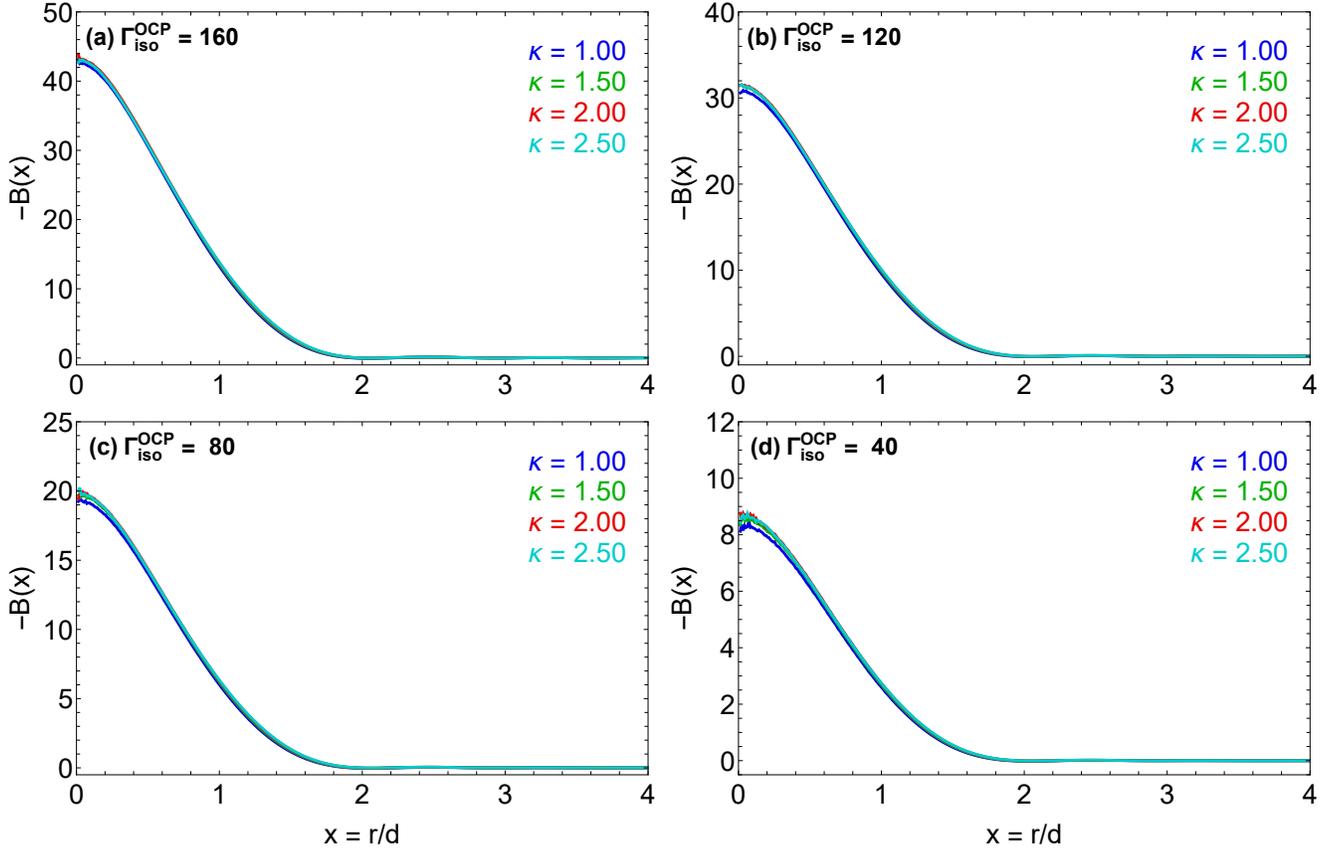}
	\caption{\enquote{Exact} bridge functions of dense YOCP liquids in the entire non-trivial range of reduced distances. (a,\,b,\,c,\,d) Results for the $4$ members ($\kappa=1.0,1.5,2.0,2.5$) of the $\Gamma_{\mathrm{ISO}}^{\mathrm{OCP}}=160,\,120,\,80,\,40$ isomorphs, respectively. Short range bridge functions $r/d\leq1.4$ are computed by applying the cavity distribution method with input from \emph{ultra-long canonical MD cavity simulations} while intermediate \& long range bridge functions $r/d\geq1.25$ are computed by applying the Ornstein-Zernike inversion method with input from \emph{ultra-accurate canonical MD simulations}. The overlapping interval $1.25\leq{r}/d\leq1.4$ is utilized to re-normalize the short range bridge functions, which are computed within an arbitrary additive constant by the cavity distribution method.}\label{fig:bridgefunction_summary}
\end{figure*}

Finally, the computed Yukawa bridge functions in the entire non-asymptotic range of $r/d\leq4$ are shown in figure \ref{fig:bridgefunction_summary} for the four isomorphic curves and all the $16$ thermodynamic state points of interest. The very high degree of the still approximate bridge function isomorph invariance is apparent in the entire range. Tabulated full range Yukawa bridge functions for these $16$ state points, organized per isomorphic curve, are available in the supplementary material\,\cite{suppleme}. Specifically, raw bridge function data are provided in the interval $0.01d<{r}\leq{5d}$ in steps of $0.01d$, \emph{i.e.} the data have been down-sampled by five given the $\Delta{r}=0.002d$ bin width employed in the histogram methods for the extraction of the cavity distribution function (inside the correlation void) and the radial distribution function (outside the correlation void). Note that the $r=0.01d$ distance is missing from the dataset for all state points, since it was judged to be poorly sampled even by the ultra-long MD cavity simulations.

\section{Summary and discussion}\label{BridgeFull}

\noindent The bridge functions of Yukawa systems were systematically computed within the correlation void in an attempt to assess the validity of the conjecture of bridge function invariance in reduced units along isomorphs, \emph{i.e.} phase diagram lines of constant excess entropy. The $16$ selected state points belong to $4$ isomorphic curves that roughly cover the entire dense liquid YOCP phase diagram up to the vicinity of the liquid-solid (bcc/fcc) phase transition. The short range bridge function was made accessible after application of the cavity distribution method with structural input from ultra-long specially designed molecular dynamics simulations that feature a tagged particle pair.

A detailed methodology was developed for the design of the tagged pair interaction potential that leads to the acquisition of approximately uniform pair statistics in the whole correlation void based on an algorithmic approach. The externally controlled tagged pair potential was decomposed into windowing and biasing components. The windowing component constrained the tagged pair within overlapping sub-intervals of the correlation void without affecting their correlations within each confinement range and was realized with the aid of two error functions. The biasing component ensured statistical uniformity within each window and was determined by successive approximations with input from cavity simulations of increasing duration, starting from a Gaussian series representation of increasing complexity and culminating into a Widom even polynomial representation.

Since poor statistics at extremely short distances are rather unavoidable with the cavity distribution method, an extrapolation technique was developed to determine the value of the bridge function at the origin. The technique was based on well-known algebraic manipulations of the Ornstein-Zernike integral equation and of the non-linear closure condition that were originally employed in the derivation of the so-called zero-separation theorems. It capitalized on the high accuracy of the Widom expansion for the cavity logarithm in order to provide a reliable extrapolation.

In order to accurately quantify the level of isomorph invariance of the Yukawa bridge functions in the short range, the effect of different uncertainties had to be analyzed. It was shown that grid, size, tail and isomorphic errors are negligible compared to the statistical errors. It was also demonstrated that there are three uncorrelated sources of statistical errors with cavity logarithm statistical uncertainties being dominant. The latter statistical errors were quantified with a block averaging procedure. The final Yukawa bridge functions, featuring error bars that stem from the total statistical uncertainties, were observed to be nearly isomorph invariant also in the short range for all the four excess entropies probed. This invariance was concluded not to be exact, since the small deviations observed between isomorphic bridge functions always exceeded the level of uncertainties.

This manuscript and its companion (Paper I, Ref.\cite{accompan}) have demonstrated that Yukawa bridge functions are isomorph invariant to a high degree in the entire non-trivial range. Hence, the conjecture of the isomorph-based empirically modified hypernetted-chain (IEMHNC) approximation has been verified for Yukawa systems rationalizing the remarkable agreement of the IEMHNC structural and thermodynamics properties with the exact results of computer simulations. It is expected with confidence that the approximate isomorph invariance of bridge functions holds for any R-simple system. However, the degree of isomorph invariance should vary between systems depending on the strength of the virial potential-energy correlations which is exceptionally high for Yukawa systems. Thus, it is important to carry out similar investigations for other R-simple liquids, \emph{e.g.} for Lennard-Jones\,\cite{outrore1} or pure exponential repulsive systems\,\cite{outrore2,outrore3}. In addition, some types of isomorph invariance concerning different quantities that are strongly linked to quasi-universal behavior such as the excess entropy scaling of transport coefficients\,\cite{outrore4,outrore5} have been observed to apply also for few systems that do not have strong enough virial potential-energy correlations\,\cite{outrore6,outrore7,outrore8} to be classified as R-simple. Therefore, it is also important to check whether an approximate isentropic invariance of bridge functions holds for systems that are not R-simple.

In spite of the fact that the bridge function constitutes one of the most important and certainly the most enigmatic static two-particle correlation function, few of its exact or even approximate properties have been so far discovered. The approximate isomorph invariance for R-simple systems can now be added to this short list. Apart from its general theoretical value and the fact that it is the building block of the successful IEMHNC approach, the isomorph invariance property has additional practical value: In general, closed-form parameterizations of the bridge function have proven to be rather elusive due to the difficulty of fitting a function of three independent variables that is anyway formidable to calculate based on first principles. To our knowledge, analytical bridge functions are currently available only for few inverse power law systems\,\cite{cavitym7,cavitym9,outrore9}, since these systems require a single independent variable for the complete specification of each thermodynamic state. The approximate property of isomorph invariance suggests that the bridge functions of R-simple systems only depend on the reduced excess entropy and the reduced distance, paving the way for future parameterization studies.

\section*{Acknowledgments}

\noindent The authors would like to acknowledge the financial support of the Swedish National Space Agency under grant no.\,143/16. This work was also partially supported by VILLUM Foundation’s Grant No.\,16515 (Matter). GPU molecular dynamics simulations were carried out at the \emph{Glass and Time} computer cluster (Roskilde University). CPU molecular dynamics simulations were carried out on resources provided by the Swedish National Infrastructure for Computing (SNIC) at the NSC (Link{\"o}ping University) partially funded by the Swedish Research Council through grant agreement no.\,2016-07213.


\begin{thebibliography}{81}



\bibitem{dustrev1} V. E. Fortov, A. V. Ivlev, S. A. Khrapak, A. G. Khrapak and G. E. Morfill, \emph{Phys. Rep.} {\bf 421}, 1 (2005)
\bibitem{dustrev2} Z. Donko, G. J. Kalman and P. Hartmann, \emph{J. Phys.: Condens. Matter} {\bf 20}, 413101 (2008)
\bibitem{dustrev3} G. E. Morfill and A. V. Ivlev, \emph{Rev. Mod. Phys.} {\bf 81}, 1353 (2009)
\bibitem{dustrev4} M. Bonitz, C. Henning and D. Block, \emph{Rep. Prog. Phys.} {\bf 73}, 066501 (2010)
\bibitem{OCPrevs1} J. L. Lebowitz and E. H. Lieb, \emph{Phys. Rev. Lett.} {\bf 22}, 631 (1969)
\bibitem{OCPrevs2} M. Baus and J. P. Hansen, \emph{Phys. Rep.} {\bf 59}, 1 (1980)
\bibitem{OCPrevs3} S. Ichimaru, H. Iyetomi and S. Tanaka, \emph{Phys. Rep.} {\bf 149}, 91 (1987)
\bibitem{dustexp1} S. A. Khrapak, B. A. Klumov, P. Huber, V. I. Molotkov \emph{et al.}, \emph{Phys. Rev. E} {\bf 85}, 066407 (2012)
\bibitem{dustexp2} H. M. Thomas, M. Schwabe, M. Y. Pustylnik, C. A. Knapek \emph{et al.}, \emph{Plasma Phys. Control. Fusion} {\bf 61} 014004 (2019)
\bibitem{dustexp3} H. Boroudjerdi, Y.-W. Kim, A. Naji, R. R. Netz \emph{et al.}, Phys. Rep. {\bf 416}, 129 (2005)
\bibitem{dustexp4} T. C. Killian, T. Pattard, T. Pohl and J. M. Rost, \emph{Phys. Rep.} {\bf 449}, 77 (2007)
\bibitem{dustexp5} K. W{\"u}nsch, J. Vorberger and D. O. Gericke, \emph{Phys. Rev. E} {\bf 79}, 010201 (2009)
\bibitem{isoYOCPg} A. A. Veldhorst, T. B. Schr{\o}der and J. C. Dyre, \emph{Phys. Plasmas} {\bf 22}, 073705 (2015)
\bibitem{isogene1} N. P. Bailey, U. R. Pedersen, N. Gnan, T. B. Schr{\o}der and J. C. Dyre, \emph{J. Chem. Phys.} {\bf 129}, 184507 (2008)
\bibitem{isogene2} T. S. Ingebrigtsen, T. B. Schr{\o}der and J. C. Dyre, \emph{Phys. Rev. X} {\bf 2}, 011011 (2012)
\bibitem{isogene3} J. C. Dyre, \emph{Phys. Rev. E} {\bf 88}, 042139 (2013)
\bibitem{isogene4} J. C. Dyre, \emph{J. Phys.: Condens. Matter} {\bf 28}, 323001 (2016)
\bibitem{isogene5} T. B. Schr{\o}der and J. C. Dyre, \emph{J. Chem. Phys.} {\bf 141}, 204502 (2014)
\bibitem{ourwork1} P. Tolias and F. Lucco Castello, \emph{Phys. Plasmas} {\bf 26}, 043703 (2019)
\bibitem{ourwork2} F. Lucco Castello, P. Tolias, J. S. Hansen, and J. C. Dyre, \emph{Phys. Plasmas} {\bf 26}, 053705 (2019)
\bibitem{ourwork3} F. Lucco Castello and P. Tolias, \emph{Contrib. Plasma Phys.} 202000105 (2020); doi:10.1002/ctpp.202000105
\bibitem{accompan} F. Lucco Castello, P. Tolias and J. C. Dyre, \emph{Testing the isomorph invariance of the bridge functions of Yukawa one-component plasmas. I. Intermediate and long range}; arXiv:2008.09078
\bibitem{isogene6} T. S. Ingebrigtsen, T. B. Schr{\o}der and J. C. Dyre, \emph{J. Phys. Chem. B} {\bf 116}, 1018 (2012)
\bibitem{isogene7} N. Gnan, T. B. Schr{\o}der, U. R. Pedersen, N. P. Bailey and J. C. Dyre, \emph{J. Chem. Phys.} {\bf 131}, 234504 (2009)
\bibitem{isogene8} J. C. Dyre, \emph{J. Phys. Chem. B} {\bf 118}, 10007 (2014)
\bibitem{isogene9} L. Costigliola, T. B. Schr{\o}der and J. C. Dyre, \emph{J. Chem. Phys.} {\bf 144}, 231101 (2016)
\bibitem{isogen10} A. K. Bacher, T. B. Sch{\o}der and J. C. Dyre, \emph{J. Chem. Phys.} {\bf 149}, 114502 (2018)
\bibitem{isogen11} L. B{\o}hling, N. P. Bailey, T. B. Schr{\o}der and J. C. Dyre, \emph{J. Chem. Phys.} {\bf 140}, 124510 (2014)
\bibitem{isomana1} U. R. Pedersen, L. Costigliola, N. P. Bailey, T. B. Schr{\o}der and J. C. Dyre, \emph{Nat. Commun.} {\bf 7}, 12386 (2016)
\bibitem{isomana2} O. Vaulina and S. A. Khrapak, \emph{J. Exp. Theor. Phys.} {\bf 90}, 287 (2000)
\bibitem{isomana3} O. Vaulina, S. Khrapak and G. Morfill, \emph{Phys. Rev. E} {\bf 66}, 016404 (2002)
\bibitem{isomana4} S. Hamaguchi, R. Farouki and D. H. E. Dubin, \emph{J. Chem. Phys.} {\bf 105}, 7641 (1996)
\bibitem{isomana5} S. Hamaguchi, R. Farouki and D. H. E. Dubin, \emph{Phys. Rev. E} {\bf 56}, 4671 (1997)
\bibitem{Hansenbo} J. P. Hansen and I. R. McDonald, \emph{Theory of Simple Liquids}, (Academic Press, London, 2006)
\bibitem{liquidbo} D. Chandler, \emph{Introduction to modern statistical mechanics}, (Oxford University Press, New York, 1987)
\bibitem{theorybo} A. Santos, \emph{A concise course on the theory of classical liquids}, (Springer, Heidelberg, 2016)
\bibitem{bridger1} C. Caccamo, \emph{Phys. Rep.} {\bf 274}, 1 (1996)
\bibitem{bridger2} J. M. Bomont, \emph{Adv. Chem. Phys.} {\bf 139}, 1 (2008)
\bibitem{bridger3} P. Attard and G. N. Patey, \emph{J. Chem. Phys.} {\bf 92}, 4970 (1990)
\bibitem{bridger4} H. L. Frisch and J. L. Lebowitz, \emph{The Equilibrium Theory of Classical Fluids}, (Benjamin, New York, 1964).
\bibitem{bridger5} J. S. Perkyns, K. M. Dyer and B. M. Pettitt, \emph{J. Chem. Phys.} {\bf 116}, 9404 (2002)
\bibitem{bridger6} S, Labik, H. Gabrielova, J. Kolafa and A. Malijevsky, \emph{Mol. Phys.} {\bf 101}, 1139 (2003)
\bibitem{bridger7} S. K. Kwak, and D. A. Kofke, \emph{J. Chem. Phys.} {\bf 122}, 104508 (2005)
\bibitem{cavitym1} G. Torrie and G. N. Patey, \emph{Mol. Phys.} {\bf 34}, 1623 (1977)
\bibitem{cavitym2} M. Llano-Restrepo and W. G. Chapman, \emph{J. Chem. Phys.} {\bf 97}, 2046 (1992)
\bibitem{cavitym3} B. Widom, \emph{J. Chem. Phys.} {\bf 39}, 2808 (1963)
\bibitem{cavitym4} D. L. Cheung, L. Anton, M. P. Allen and A. J. Masters, \emph{Phys. Rev. E} {\bf 73}, 061204 (2006)
\bibitem{cavitym5} L. Belloni, \emph{J. Chem. Phys.} {\bf 147}, 164121 (2017)
\bibitem{cavitym6} D. Tomazic, F. Hoffgaard and S. M. Kast, \emph{Chem. Phys. Lett.} {\bf 591}, 237 (2014)
\bibitem{RUMDref1} N. P. Bailey, T. S. Ingebrigtsen, J. S. Hansen, A. A. Veldhorst \emph{et al.}, \emph{SciPost Phys.} {\bf 3}, 038 (2017)
\bibitem{RUMDref2} S. Plimpton, \emph{J. Comp. Phys} {\bf 117}, 1 (1995)
\bibitem{RUMDref3} S. Toxvaerd and J. C. Dyre, \emph{J. Chem. Phys.} {\bf 134}, 081102 (2011)
\bibitem{cavitym7} M. Llano-Restrepo and W. G. Chapman, \emph{J. Chem. Phys.} {\bf 100}, 5139 (1994)
\bibitem{cavitym8} J. M. Caillol and D. Gilles, \emph{J. Phys. A: Math. Gen.} {\bf 36}, 6243 (2003)
\bibitem{cavitym9} A. Malijevsky and S. Labik, \emph{Mol. Plys.} {\bf 60}, 663 (1987)
\bibitem{SMSAmina} P. Tolias, S. Ratynskaia and U. de Angelis, \emph{Phys. Rev. E} {\bf 90}, 053101 (2014)
\bibitem{SMSAminb} P. Tolias, S. Ratynskaia and U. de Angelis, \emph{Phys. Plasmas} {\bf 22}, 083703 (2015)
\bibitem{zerosep1} Y. Rosenfeld, \emph{Phys. Rev. A} {\bf 24}, 2805 (1981)
\bibitem{zerosep2} L. L. Lee and K. S. Shing, \emph{J. Chem. Phys.} {\bf 91}, 477 (1989)
\bibitem{zerosep3} L. L. Lee, \emph{J. Chem. Phys.} {\bf 103}, 9388 (1995)
\bibitem{hypervir} M. P. Allen and D. J. Tildesley, \emph{Computer simulation of liquids}, (Clarendon Press, Oxford, 1989)
\bibitem{decompo1} Y. Rosenfeld and P. Tarazona, \emph{Mol. Phys.} {\bf 95}, 141 (1998)
\bibitem{decompo2} Y. Rosenfeld, \emph{Phys. Rev. E} {\bf 62}, 7524 (2000)
\bibitem{zerosep4} J. R. Henderson, \emph{Mol. Phys.} {\bf 48}, 389 (1983)
\bibitem{sizeexp1} J. L. Lebowitz and J. K. Percus, \emph{Phys. Rev.} {\bf 122}, 1675 (1961)
\bibitem{sizeexp2} J. L. Lebowitz, J. K. Percus and L. Verlet, \emph{Phys. Rev.} {\bf 153}, 250 (1967)
\bibitem{cavityn1} S. Ogata, \emph{Phys. Rev. E} {\bf 53}, 1094 (1996)
\bibitem{cavityn2} R. Fantoni and G. Pastore, \emph{J. Chem. Phys.} {\bf 120}, 10681 (2004)
\bibitem{suppleme} See [URL link-with-supplementary-material] for extensive tabulations of the raw data for the Yukawa bridge functions of the $16$ state points of interest (equally distributed amongst four isomorphic curves) in the entire non-asymptotic range.
\bibitem{outrore1} T. B. Schr{\o}der, N. Gnan, U. R. Pedersen, N. P. Bailey and J. C. Dyre, \emph{J. Chem. Phys.} {\bf 134}, 164505 (2011)
\bibitem{outrore2} A. K. Bacher, T. B. Schr{\o}der and J. C. Dyre, \emph{J. Chem. Phys.} {\bf 149}, 114501 (2018)
\bibitem{outrore3} A. K. Bacher, T. B. Schr{\o}der and J. C. Dyre, \emph{J. Chem. Phys.} {\bf 149}, 114502 (2018)
\bibitem{outrore4} Y. Rosenfeld, \emph{Phys. Rev. A} {\bf 15}, 2545 (1977)
\bibitem{outrore5} J. C. Dyre, \emph{J. Chem. Phys.} {\bf 149}, 210901 (2018)
\bibitem{outrore6} R. Errington, T. M. Truskett and J. Mittal, \emph{J. Chem. Phys.} {\bf 125}, 244502 (2006)
\bibitem{outrore7} W. P. Krekelberg, T. Kumar, J. Mittal, J. R. Errington and T. M. Truskett, \emph{Phys. Rev. E} {\bf 79}, 031203 (2009)
\bibitem{outrore8} M. Agarwal, M. Singh, R. Sharma, M. P. Alam and C. Chakravarty, \emph{J. Phys. Chem. B} {\bf 114}, 6995 (2010)
\bibitem{outrore9} H. Iyetomi, S. Ogata and S. Ichimaru, \emph{Phys. Rev. A} {\bf 46}, 1051 (1992)
\end{thebibliography}
\end{document}